\documentclass[prd,tightenlines,nofootinbib,superscriptaddress]{revtex4-2}
\usepackage{amsfonts,amsmath,amssymb,amsthm,bbm,hyperref}
\usepackage{graphicx}
\usepackage{booktabs}
\usepackage{array}

\DeclareSymbolFont{yhlargesymbols}{OMX}{yhex}{m}{n}
\DeclareMathAccent{\widetriangle}{\mathord}{yhlargesymbols}{"E6}

\newcommand{\C}{{\mathbb C}}

\newcommand{\cC}{{\mathcal C}}

\newcommand{\cX}{{\mathcal X}}
\newcommand{\cY}{{\mathcal Y}}
\newcommand{\SU}{\mathrm{SU}}
\newcommand{\SL}{\mathrm{SL}}

\newcommand{\U}{\mathrm{U}}

\newcommand{\w}{\wedge}
\newcommand{\id}{\mathbb{I}}

\newcommand{\be}{\begin{equation}}
\newcommand{\ee}{\end{equation}}
\newcommand{\beq}{\begin{eqnarray}}
\newcommand{\eeq}{\end{eqnarray}}
\newcommand{\bea}{\begin{eqnarray}}
\newcommand{\eea}{\end{eqnarray}}

\newcommand{\su}{{\mathfrak su}}

\newcommand{\la}{\langle}
\newcommand{\ra}{\rangle}

\newcommand{\tr}{{\mathrm Tr}}
\newcommand{\f}{\frac}

\newcommand{\vphi}{\varphi}
\def\eps{\epsilon}

\newcommand{\rd}{\mathrm{d}}

\newcommand{\bz}{\overline{z}}

\def\pp{\partial}

\def\tH{\widetilde{H}}

\def\vphi{\varphi}

\def\bz{\bar{z}}
\def\bF{\bar{F}}

\def\vX{\vec {X}}
\def \hu{\hat{u}}
\def\bE{\bar{E}}
\def\bF{\bar{F}}
\def\vcX{\vec{\cX}}
\def\vcY{\vec{\cY}}

\def\tz{\tilde{z}}
\def\vsigma{\vec{\sigma}}

\newcommand{\mat}[2]{\left(\begin{array}{#1}#2
\end{array}\right)}

\begin{document}

\title{Classical Dynamics for Loop Gravity: The 2-Vertex Model}

\author{Eneko Aranguren}\email{enekoarangurenruiz@gmail.com}
\affiliation{Department of Physics, University of the Basque Country UPV/EHU, P.O. Box 644, 48080 Bilbao, Spain}
\author{Iñaki Garay}\email{inaki.garay@ehu.eus}
\affiliation{Department of Physics, University of the Basque Country UPV/EHU, P.O. Box 644, 48080 Bilbao, Spain}
\affiliation{EHU Quantum Center, University of the Basque Country UPV/EHU, Leioa, Spain}
\author{Etera R. Livine}\email{etera.livine@ens-lyon.fr}
\affiliation{Univ Lyon, ENS de Lyon, CNRS, Laboratoire de Physique, UMR 5672, 69007 Lyon, France}

\date{\small \today}

\begin{abstract}
The study of toy models in loop quantum gravity (LQG), defined as truncations of the full theory, is relevant to both the development of the LQG phenomenology, in cosmology and astrophysics, and the progress towards the resolution of the open issues of the theory, in particular the implementation of the dynamics. Here, we study the dynamics of spin network states of quantum geometry defined on the family of graphs consisting in 2 vertices linked by an arbitrary number of edges, or {\it 2-vertex model} in short. A symmetry reduced sector of this model --to isotropic and homogeneous geometries-- was successfully studied in the past, where interesting cosmological insights were found.
We now study the evolution of the classical trajectories for this system in the general case, for arbitrary number of edges with random initial configurations.
We use the spinorial formalism and its clear interpretation of spin networks in terms of discrete twisted geometries, with the quantum 3d space made of superpositions of polyhedra glued together by faces of equal area.
Remarkably, oscillatory and divergent regimes are found with a universal dependence on the coupling constants of the Hamiltonian and independent of the initial spinors or the number of edges. Furthermore, we explore the evolution of the associated polyhedra as well as their volumes and areas.
\end{abstract}

\maketitle

\tableofcontents

\section*{Introduction}

Loop quantum gravity (LQG) offers a description of the Planck scale geometry of space-time. Its dynamics, in its semi-classical regime, is meant to reproduce the Hamiltonian evolution of general relativity (possibly with corrections in both the infrared and ultraviolet sectors). Analyzing the coarse-graining and renormalization flow of the full theory over the several orders of magnitude bridging between the Planck scale and astrophysical scales is an arduous task. From this perspective, studying consistent truncated models turns out to be extremely useful in gaining intuition about the whole theory, exploring its complex phase diagram and investigating its cosmological and astrophysical phenomenology.

The simplest family of truncations that one may explore is given by the 2-vertex model that was studied in the past with applications to cosmology and black hole modeling \cite{Rovelli:2008aa,Borja:2010gn,Borja:2010rc,Borja:2011ha,Livine:2011up}. The model consists in truncating the LQG dynamics to spin network states living on a graph made of two vertices linked by an arbitrary number $N$ of edges.
Despite the simplicity of the model and its straightforward cosmological interpretation, this model has been superficially studied, up to now, in a classical or semi-classical setting, with the strong assumption of isotropy and homogeneity, which amounts to focusing on its symmetry reduction imposing a global $\U(N)$ symmetry. In this paper, we study the dynamics of the model for general inhomogeneous configurations. Investigating the whole space of configurations of the model is progress both towards  the study of dynamics of non-homogeneous spin network states in LQG and towards the inclusion of anisotropies in the cosmological application of the 2-vertex model.

\smallskip

If one truncates LQG to dynamics on a fixed graph, one works on the corresponding holonomy-flux phase space, consisting in a collection of a certain number of particles living on the Lie group $\SU(2)$ (one particle per graph edge) together with constraints imposing the $\SU(2)$ gauge invariance. Upon quantization, this leads to the Hilbert space of spin network states with support on that graph. Now, if one is interested in Hamiltonian dynamics, the  evolution generated by the classical Hamiltonian on the phase space gives the leading order of the quantum dynamics, at least for coherent wave-packets. We  thus start by studying classical Hamiltonian dynamics on 2-vertex graphs.

We use the spinorial parametrization of the loop gravity phase space \cite{Livine:2011gp,Livine:2013zha,Dupuis:2011dyz,Borja:2010rc}. One starts with spinor variables, as complex 2-vectors, dressing the graph, from which one can build back both holonomy and flux variables. The spinors have canonical Poisson brackets and can be interpreted as Darboux coordinates for the phase space on a fixed graph. These spinors are an outstanding tool for LQG and spinfoams. At the mathematical level, they are the natural variables for $\SU(2)$ 
unitary representations, whose Hilbert space simply consists in polynomial wave-functions in the spinors (e.g. \cite{Livine:2011gp,Alesci:2016dqx}). At the geometrical level, the spinors provide a simple parametrization of twisted geometries \cite{Freidel:2010bw} and clarify the structure and solution to the simplicity constraints, which allow to derive general relativity from topological BF theory \cite{Dupuis:2010iq,Dupuis:2011fz,Dupuis:2011wy,Speziale:2012nu,Dona:2020xzv}.
These two key properties of the spinorial reformulation of the loop gravity phase space has led to several interesting results:
a definition of coherent spin network states (e.g. \cite{Dupuis:2010iq,Dupuis:2011fz}), subsequently their application to the definition of EPRL-type spinfoam models (e.g. \cite{Livine:2007ya}) and the semi-classical analysis of the corresponding spinfoam amplitudes in the large spin asymptotic limit \cite{Barrett:2009mw,Barrett:2010ex,Han:2011rf,Han:2011re,Han:2013gna,Dona:2020yao}, the analysis of the space of interwiners as the quantization of classical polyhedra and the discovery of their fine structure in terms of $\U(N)$ representations \cite{Freidel:2009ck,Freidel:2010tt,Livine:2012cv,Livine:2013tsa}, the derivation of the Hamiltonian constraints from spinfoam amplitude recursion relations \cite{Bonzom:2011hm,Bonzom:2011nv}, progress on the coarse-graining flow of EPRL amplitudes \cite{Banburski:2014cwa}, extension to $q$-deformed LQG for spin networks with non-vanishing cosmological constant \cite{Bonzom:2014wva,Dupuis:2014fya,Bonzom:2021yma}, and insights into the renormalization flow of LQG and cosmology \cite{Bodendorfer:2020nku,Bodendorfer:2020ovt}.

Concretely, given a graph, one dresses every half-edge of the graph with a spinor variable in $\C^2$. One further imposes two sets of constraints: matching constraints on every edge and closure constraints on every vertices.
Around a vertex, one projects the spinors onto real 3-vectors, thereby deriving the fluxes, then Minkowski's theorem on polytopes \cite{Minkowski1897} ensures that one can reconstruct a polyhedron of $N$ faces from any collection of $N$ spinors satisfying the closure constraint, such that the fluxes give the normal vectors to the polyhedron's faces.
Along an edge, the mapping between the two spinors, living on the two halves of the edge, define the $\SU(2)$ holonomy or transport along the edge. The matching constraint on the edge imposes that the classical polyhedra reconstructed at the vertices of the edge are glued by faces of equal area. Such discrete geometry  are called twisted geometries \cite{Freidel:2010aq}. Upon quantization, twisted geometries become spin network states, closure constraints lead to the $\SU(2)$ invariance of intertwiners at the vertices, matching constraints ensure that each edge carries a $\SU(2)$ spin representing a $\SU(2)$ holonomy.

\smallskip

We apply this machinery to the 2-vertex graph and study the classical dynamics of spinor networks and twisted geometries. More precisely, we analyze the classical trajectories for different number of edges and Hamiltonian normalization of the 2-vertex model, as well as for the generic case given by an arbitrary number of edges with random initial configurations of spinors. Furthermore, using the reconstruction algorithm by Sellaroli \cite{Sellaroli:2017wwc}, we reconstruct the polyhedra and compute the evolution of their volume and shape.
Systematically studying a large number of initial conditions for various values of the coupling constants entering the Hamiltonian, we identify a universal behavior of the regimes (oscillatory or divergent) depending on the coupling constants (and with no dependence on the initial conditions). Remarkably, this same dependence was found for the reduced isotropic and homogeneous case at both classical level \cite{Borja:2010rc}  and quantum level \cite{Borja:2010gn}. In light of the interpretation of this symmetry-reduced regime in terms of homogeneous loop quantum cosmology (LQC),
this opens the way to extrapolate this analogy to the general case treated in the present paper.

The paper is structured in the following way. In section \ref{sec:2-vertex}, we  review the spinor formalism applied to the 2-vertex graph and we  discuss different general Hamiltonians for the model (commuting with the constraints) that may be proposed. In section \ref{sec:2vertexmodel}, the simplest case of only $2$ edges will be studied, which will be useful in order to explore the different regimes found depending on the coupling constants and to gain insight into the dynamics of the 2-vertex model. The case of the isosceles tetrahedron explored in section \ref{sec:tetrahedron} is very important, given that the tetrahedron represents the simplest polyhedron (corresponding to a 4-valent intertwiner in the LQG theory) with non-trivial volume. Finally, section \ref{sec:four} presents the general case for an arbitrary number of edges and random initial conditions.

\section{The 2-vertex model}
\label{sec:2-vertex}

\begin{figure}[htb]
\begin{center}
\includegraphics[width=0.45\textwidth]{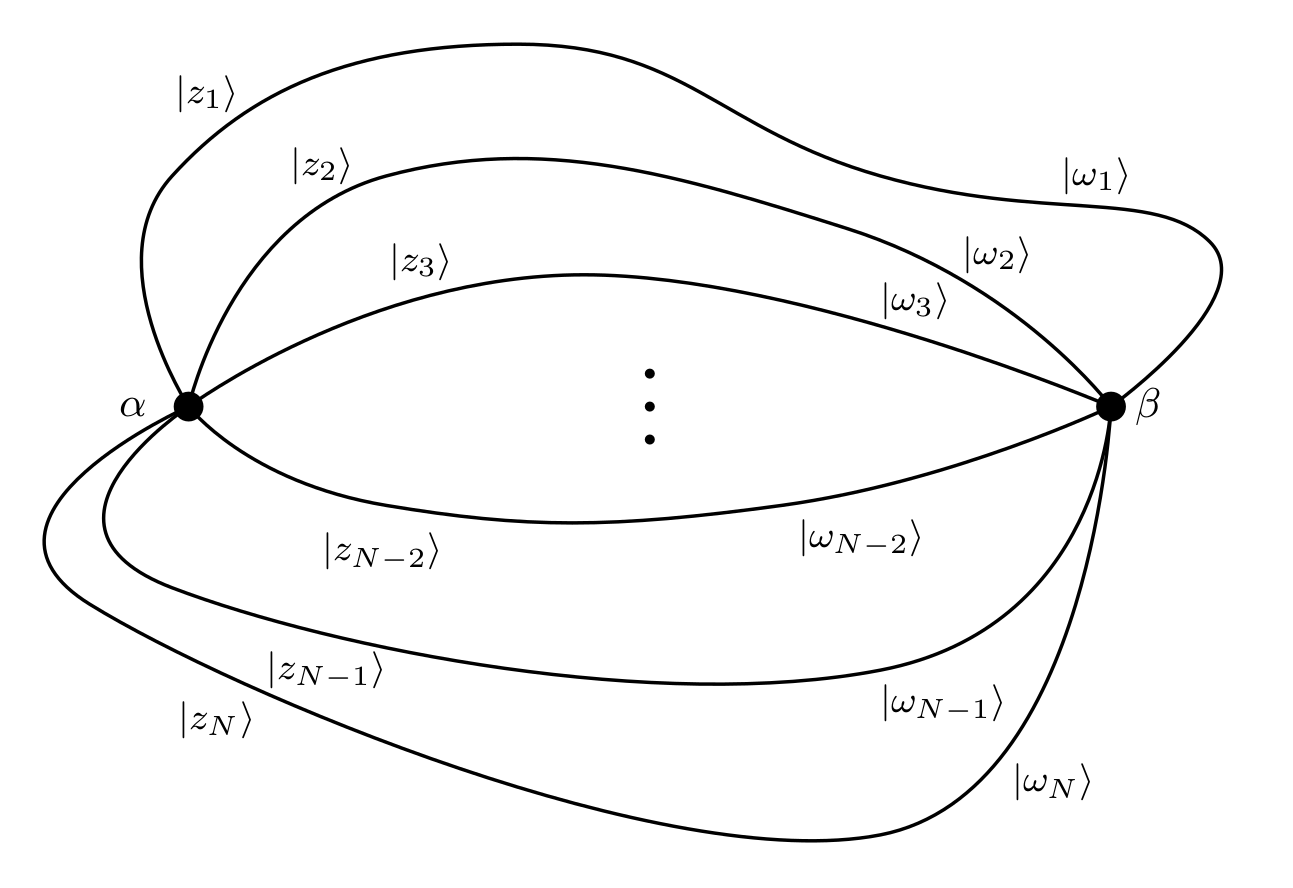}
\caption{The 2-vertex graph (also considered in \cite{Borja:2010gn,Borja:2010rc}) that consists of 2 vertices labeled with $\alpha$ and $\beta$ and an arbitrary number of $N$ edges endowed with a couple of spinors $|z_i\ra$ and $|\omega_i\ra$ attached to each of them.}
\label{fig:2vertexfig}
\end{center}
\end{figure}

We consider the same family of models studied in \cite{Borja:2010gn,Borja:2010rc}, given by a graph with two vertices, noted $\alpha$ and $\beta$ and an arbitrary number of edges $N$ between them (see figure \ref{fig:2vertexfig}). That previous work focused however on the dynamics for a reduced sector, considering only states invariant under a global $\U(N)$ symmetry. This led to a single state for each value of the sum of the spins carried by the edges, with no local freedom for the intertwiners at the vertices or the individual spins on the edges.
In the present paper, we push the analysis and explore the general case, for arbitrary spin fluctuations and intertwiner data.

\subsection{Spinor phase space on the 2-vertex graph}

Following the spinor formalism for LQG \cite{Livine:2011gp,Livine:2013zha,Dupuis:2011dyz,Borja:2010rc}, 
we  define a spinor $|z\ra \in \mathbb{C}^2$, its conjugate $\la z|\in \mathbb{C}^2$ and the spinor $|z] \in \mathbb{C}^2$:
\be
|z\ra=\mat{c}{z^0\\z^1}, \qquad
\la z|=\mat{cc}{\bar{z}^0 &\bar{z}^1},\qquad |z]:=-i\sigma^2|\bar{z}\ra, \quad \text{with}\quad\sigma^2=\mat{cc}{0 & -i\\ i & 0}.
\ee
Therefore, for the 2-vertex graph, we  attach $N$ spinor variables at every vertex, $|z^{\alpha}_{k}\ra$ and $|z^{\beta}_{k}\ra$, with $k$ running from 1 to $N$ (one spinor for each edge and vertex). In order to ease the notation (and avoid multiple indices), we will write $|z^{\alpha}_{k}\ra=|z_{k}\ra$ and $|z^{\beta}_{k}\ra=|w_{k}\ra$.

Each spinor variable is provided with canonical Poisson bracket
\be
\{z^{A},z^{B}\}=\{\bz^{A},\bz^{B}\}=0,\qquad \{z^{A},\bz^{B}\}=-i\delta^{AB}\,, \quad
\text{with}\quad A, B=0,1\,;
\ee
and it defines a 3d real vector by its projection onto the Pauli matrices, normalized to $(\sigma^{a})^{2}=\id$, for each $a=1,2,3$:
\be
X^{a}=\f12\la z | \sigma^{a}|z\ra={\f12}\sigma^{a}_{AB}\bz_{A}z_{B},\qquad
X\equiv |\vX|=\f12\la z |z\ra\,.
\ee
These vector components naturally form a $\su(2)$ Lie algebra,
\be
\{X,X^{a}\}=0\,,\qquad
\{X^{a},X^{b}\}
=\eps^{abc}X^{c}
\,,
\ee
whose Casimir is obviously the squared norm, $\vX^{2}=X^{2}$, and who generates the $\SU(2)$ group action in the spinor,
\be
e^{{\{\theta \hu\cdot\vX, \bullet\}}}\,|z\ra
=
g(\theta,\hu)\,\,|z\ra
\,,
\qquad
g(\theta,\hu)=e^{i\f\theta2\,\hu\cdot\vec{\sigma}}
\in\SU(2)\,.
\ee

Along each edge linking the two vertices, we define a $\SU(2)$ group element mapping the spinors around $\alpha$ and the spinors around $\beta$:
\be
g_{k}=\f{|w_{k}]\la z_{k}|-|w_{k}\ra [z_{k}|}{\sqrt{\la z_{k} |z_{k}\ra\la w_{k} |w_{k}\ra}}\in\SU(2)
\,,\qquad
g_{k}\,\f{|z_{k}\ra}{\sqrt{\la z_{k} |z_{k}\ra}}=\f{|w_{k}]}{\sqrt{\la  w_{k} |w_{k}\ra}}
\,,\quad
g_{k}\,\f{|z_{k}]}{\sqrt{\la z_{k} |z_{k}\ra}}=\,-\f{|w_{k}\ra}{\sqrt{\la  w_{k} |w_{k}\ra}}
\,.
\ee
Let us call $X^{a}_{k}$ the 3-vectors defined by the spinors $|z_{k}\ra$ around the vertex $\alpha$ and $Y^{a}_{k}$ the 3-vectors defined by the spinors $|w_{k}\ra$ around the vertex $\beta$. Then, if we assume norm-matching condition along each edge,
\be
\cC_{k}=\la z_{k} |z_{k}\ra-\la w_{k} |w_{k}\ra=0\,,
\ee
$X_{k}$, $Y_{k}$ and $g_{k}$ form a $T^{*}\SU(2)$ Lie algebra:
\be
\{g_{k},g_{k}\}\sim0
\,,\qquad
\{X_{k}^{a},g_{k}\}=g_{k}\,\left(\f{-i\sigma^{a}}{2}\right)
\,,\qquad
\{Y_{k}^{a},g_{k}\}=\left(\f{+i\sigma^{a}}{2}\right)\,g_{k}
\,,\qquad
\{X_{k},Y_{k}\}=0.
\ee

\subsection{LQG Hamiltonian}

In order to define the LQG Hamiltonian driving the time evolution of the data living on the 2-vertex graph, we introduce $\SU(2)$ invariant observables around each of the two vertices. Following \cite{Borja:2010rc,Livine:2011gp}, we introduce:
\be
E_{kl}^{\alpha}=\la z_{k}^{\alpha}|z_{l}^{\alpha}\ra
\,,\qquad
F_{kl}^{\alpha}=[ z_{k}^{\alpha}|z_{l}^{\alpha}\ra
\,,\qquad
\bF_{kl}^{\alpha}=\la z_{l}^{\alpha}|z_{k}^{\alpha}]
=-\la z_{k}^{\alpha}|z_{l}^{\alpha}]\,,
\ee
and similarly for $\beta$. They satisfy the following symmetry and reality conditions:
\be
\bE_{kl}=E_{lk}\,,\qquad
F_{kl}=-F_{lk}\,.
\ee
On top of this, the observables $F_{ij}$ also satisfy the Pl\"ucker relations:
{ 
\be
F_{ij}F_{kl}=F_{il}F_{kj}+F_{ik}F_{jl}.
\ee
}

Both observables $E_{ij}$ and $F_{ij}$ are invariant under global $\SU(2)$ transformations of the spinors around $\alpha$,
\be
|z_{k}^{\alpha}\ra \overset{h\in\SU(2)}{\mapsto} h\,|z_{k}^{\alpha}\ra
\,,\qquad
E_{kl}^{\alpha}\mapsto E_{kl}^{\alpha}
\,,\qquad
F_{kl}^{\alpha}\mapsto F_{kl}^{\alpha}\,.
\ee
Since this $\SU(2)$ action is generated by the Poisson bracket with the total vector (or closure vector) around $\alpha$, $\vcX=\sum_{k}\vX_{k}$, this $\SU(2)$-invariance simply corresponds to a vanishing Poisson bracket:
\be
\{\cX^{a},E^{\alpha}_{kl}\}=\{\cX^{a},F^{\alpha}_{kl}\}=0\,,
\ee
and similarly around the vertex $\beta$, for which we introduce the total vector $\vcY$.

We can thus introduce a Hamiltonian coupling the geometry around both vertices:
\be
H=\sum_{k,l}\lambda E^{\alpha}_{kl}E^{\beta}_{kl}
+\gamma  F^{\alpha}_{kl}F^{\beta}_{kl}
+\bar{\gamma} \bF^{\alpha}_{kl}\bF^{\beta}_{kl}
\,.
\label{eq:LQGham}
\ee
This ansatz has two essential properties:
\begin{itemize}
\item it is invariant under $\SU(2)$ (gauge) transformations acting independently on each vertex, which translates into the following vanishing Poisson brackets,
\be
\{H,\vcX\}=\{H,\vcY\}=0\,,
\ee
\item it is invariant under $\U(1)$ phase transformations along every edge, which translates into the following vanishing Poisson brackets,
\be
\{H,\cC_{k}\}=0,\quad\forall k
\,.
\ee
\end{itemize}
This means that $\vcX$, $\vcY$ and the $\cC_{k}$ are all constants of motion. In particular, if we choose closed and matching initial conditions, i.e. $\vcX=\vcY=0$ and $\cC_{k}=0$ for all $k$'s, then the configuration of spinors remains closed and matching during the time evolution. Notice that the closure constraint (given by $\vcX=\vcY=0$) would translate, after quantization, into the $\SU(2)$ invariance of the intertwiner \cite{Livine:2011gp,Borja:2010rc}. Furthermore, considering the Minkowski theorem \cite{Minkowski1897}, closed configurations correspond uniquely (up to translations and rotations) to polyhedra whose face normals are given by $\vec{X}_i$ and whose areas of the faces are given by their moduli $X_i$. 

We may compute easily the equations of motion for the Hamiltonian \eqref{eq:LQGham}:
\be
\rd_{t}|z_{i}\ra=-\{H,|z_{i}\ra\}=\sum_j\left(2i\bar{\gamma}\bar{F}_{ij}^{\beta}|z_j]-i\lambda E_{ij}^{\beta}|z_j\ra\right),
\label{eomLQGham}
\ee
and similarly for the spinors $|\omega_i\ra$.

On the other hand, if we define for the Hamiltonian terms the notation
\be 
e_0=\sum_{k,l}E^{\alpha}_{kl}E^{\beta}_{kl},\quad f_0=\sum_{k,l}F^{\alpha}_{kl}F^{\beta}_{kl},
\quad \bar{f}_0=\sum_{k,l}\bF^{\alpha}_{kl}\bF^{\beta}_{kl},
\ee
the following brackets are satisfied (using the closure and matching constraints):
\be
\{e_0,f_0\}=2iEf_0,\qquad
\{e_0,\bar{f}_0\}=-2iE\bar{f}_0,\qquad
\{f_0,\bar{f}_0\}=-4iEe_0,
\ee
with $E=E^\alpha=E^\beta=\sum_i\la z_i|z_i\ra$.

\subsection{Lattice gauge theory}

It is extremely useful to understand the Hamiltonian ansatz \eqref{eq:LQGham} in terms of $\SU(2)$ holonomy and lattice gauge theory. For instance, it is very natural to consider the $\SU(2)$ holonomy for a loop linking the two vertices along a pair of edges $(k,l)$:
\be
\chi_{kl}=\tr g_{k}g_{l}^{-1}=
\f{\tr(|w_{k}]\la z_{k}|-|w_{k}\ra [z_{k}|)(|z_{l}\ra [w_{l}|-|z_{l}]\la w_{l}|)}
{\sqrt{\la z_{k} |z_{k}\ra\la w_{k} |w_{k}\ra\la z_{l} |z_{l}\ra\la w_{l} |w_{l}\ra}}
=
\f{ E^{\alpha}_{kl}E^{\beta}_{kl} +E^{\alpha}_{lk}E^{\beta}_{lk}+F^{\alpha}_{kl}F^{\beta}_{kl}+ \bF^{\alpha}_{kl}\bF^{\beta}_{kl}
}
{4\sqrt{X_{k}X_{l}Y_{k}Y_{l}}}
\,.
\label{eq:LGTHam}
\ee
Summing over all pairs of links with equal weights gives a lattice gauge theory Hamiltonian:
\be
\tH_{0}=\sum_{k,l}\chi_{kl}
=\sum_{k,l}
\f{ E^{\alpha}_{kl}E^{\beta}_{kl} +E^{\alpha}_{lk}E^{\beta}_{lk}+F^{\alpha}_{kl}F^{\beta}_{kl}+ \bF^{\alpha}_{kl}\bF^{\beta}_{kl}
}
{4\sqrt{X_{k}X_{l}Y_{k}Y_{l}}}
=e+f+\bar{f}\,,
\ee
with the more compact notations:
\be
e=\f12\sum_{k,l}
\f{ E^{\alpha}_{kl}E^{\beta}_{kl}}
{\sqrt{X_{k}X_{l}Y_{k}Y_{l}}}
\,,\qquad
f=\f14\sum_{k,l}
\f{ F^{\alpha}_{kl}F^{\beta}_{kl}}
{\sqrt{X_{k}X_{l}Y_{k}Y_{l}}}
\,,\qquad
\bar{f}=\f14\sum_{k,l}
\f{ \bar{F}^{\alpha}_{kl}\bar{F}^{\beta}_{kl}}
{\sqrt{X_{k}X_{l}Y_{k}Y_{l}}}
\,.
\ee
It was actually shown in \cite{Borja:2010rc,Livine:2011up} that the LQG version of the Hamiltonian can be written as the lattice gauge theory Hamiltonian with two flux-vector insertions, with terms of the type $\tr X_{k}X_{l}g_{k}g_{l}^{-1}$.
There are two key differences:
\begin{itemize}
\item this version of the Hamiltonian has a fixed ratio between the $E$ coupling and the $F$ coupling. This necessarily peaks the energy around flat configurations, which corresponds to flat cosmology in the context of (loop) (quantum) cosmology \cite{Livine:2011up}. We can generalize this Hamiltonian allowing for arbitrary couplings in front of $e$ and $f$, which simply modify the LQG Hamiltonian by norm factors:
\be
\tH
=
\sum_{k,l}\f1{\sqrt{X_{k}X_{l}Y_{k}Y_{l}}}\Big{[}\lambda E^{\alpha}_{kl}E^{\beta}_{kl}
+\gamma  F^{\alpha}_{kl}F^{\beta}_{kl}
+\bar{\gamma} \bF^{\alpha}_{kl}\bF^{\beta}_{kl}\Big{]}
\,.
\ee

\item the norm factors lead to a highly non-trivial modification of the Lie brackets between Hamiltonian terms, and we have not obtained a closed algebra anymore.

\end{itemize}

\subsection{The intermediate normalization}

We may also consider an intermediate normalization given by:
\be
H_\text{inter}
=
\f{2}{\sqrt{E^{\alpha} E^{\beta}}}\sum_{k,l}\Big{[}\lambda E^{\alpha}_{kl}E^{\beta}_{kl}
+\gamma  F^{\alpha}_{kl}F^{\beta}_{kl}
+\bar{\gamma} \bF^{\alpha}_{kl}\bF^{\beta}_{kl}\Big{]}
\,.
\label{IntermediateHam}
\ee
In this case, we may define
\be 
e_{1}=\f{2}{\sqrt{E^{\alpha} E^{\beta}}}e_0\,,\quad
f_{1}=\f{2}{\sqrt{E^{\alpha} E^{\beta}}}f_0\,,\quad
\bar{f}_{1}=\f{2}{\sqrt{E^{\alpha} E^{\beta}}}\bar{f}_0\,,
\ee
that satisfy the following Poisson brackets:
\beq
\{e_{1},f_{1}\}&=& 4i\left(1-\f{e_1}{2E}\right)f_1\,,\\
\{e_1,\bar{f}_1\}&=& -4i\left(1-\f{e_1}{2E}\right)\bar{f}_1\,,\\
\{f_1,\bar{f}_1\}&=&-4i\left(2e_1-\f{f_1\bar{f}_1}{E}\right)\,.
\eeq

\section{The 2 vertex model with 2 edges}
\label{sec:2vertexmodel}

\begin{figure}[htb]
\begin{center}
\includegraphics[width=0.4\textwidth]{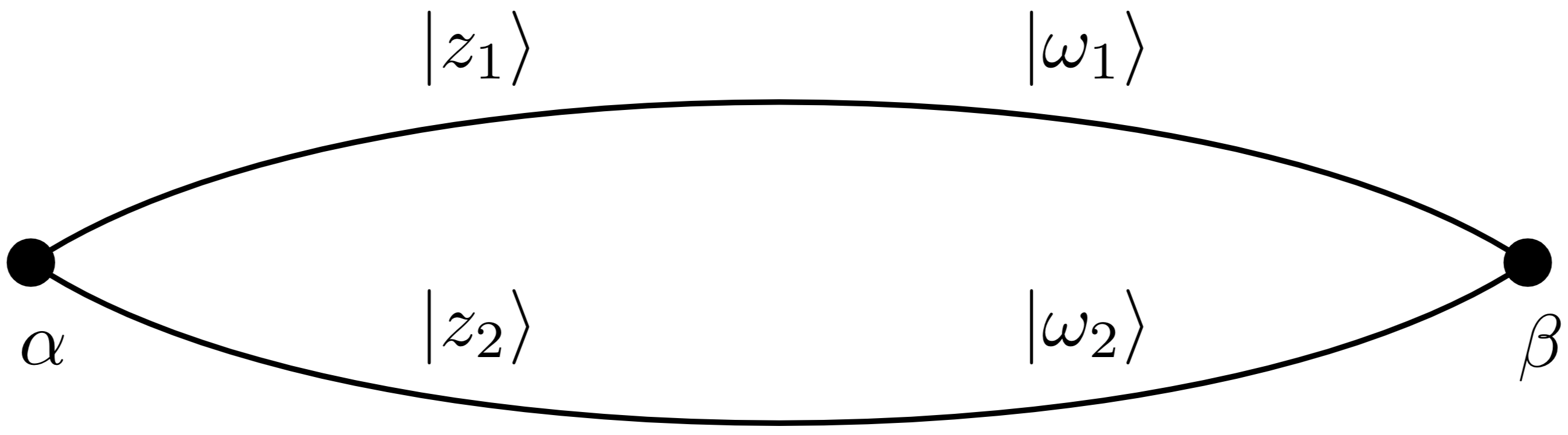}
\caption{The 2-vertex graph with only 2 edges and with spinors given by equation \eqref{eq:spinors2edges}.}
\label{fig:2edges}
\end{center}
\end{figure}

We will consider now the simplest non-trivial 2-vertex graph, with only $N=2$ edges (see figure \ref{fig:2edges}). The relations between the spinors that satisfy the closure constraint are the following:

\begin{equation}
\begin{aligned}
    &|z_{1}\rangle=|z\rangle,\qquad \qquad &|\omega_{1}\rangle &=|\omega\rangle,\\
    &|z_{2}\rangle=e^{i\theta}|z],&|\omega_{2}\rangle &=e^{i\varphi}|\omega].
\end{aligned}
  \label{eq:spinors2edges}
\end{equation}

In this section we will study the evolution of the model with 2 edges under the dynamics given by the three Hamiltonians considered in the previous section (section \ref{sec:2-vertex}).

\subsection{Loop Gravity Dynamics}\label{subsec:2edgesHLQG}

The LQG Hamiltonian \eqref{eq:LQGham} for this simple case of 2 edges may be explicitly written as:
\be
\begin{aligned}
H=\,&\lambda(\la z_{1}|z_{1}\ra\la w_{1}|w_{1}\ra+\la z_{2}|z_{2}\ra\la w_{2}|w_{2}\ra)\\
+\,&\lambda(\la z_{1}|z_{2}\ra\la w_{1}|w_{2}\ra+\la z_{2}|z_{1}\ra\la w_{2}|w_{1}\ra)\\
+\,&2\gamma[ z_{1}|z_{2}\ra[ w_{1}|w_{2}\ra+2\bar{\gamma}\la z_{1}|z_{2}]\la w_{1}|w_{2}]
\,,
\label{eq:LQGHam_2edges}
\end{aligned}
\ee
with the corresponding evolution equations:
\beq
\rd_{t}|z_{1}\ra&=-\{H,|z_{1}\ra\}
&= -i\lambda\la w_{1}|w_{1}\ra\,|z_{1}\ra-i\lambda\la w_{1}|w_{2}\ra\,|z_{2}\ra
-2i\bar{\gamma}\la w_{1}|w_{2}]\,|z_{2}]\,,\\
\rd_{t}|w_{1}\ra&=-\{H,|w_{1}\ra\} 
&=
-i\lambda\la z_{1}|z_{1}\ra\,|w_{1}\ra-i\lambda\la z_{1}|z_{2}\ra\,|w_{2}\ra
-2i\bar{\gamma}\la z_{1}|z_{2}]\,|w_{2}]
\,,
\eeq
and similarly for the spinors $z_{2}$ and $w_{2}$.
Applied to closed initial conditions (equations \ref{eq:spinors2edges}),
where the phases $\theta$ and $\vphi$ should a priori be allowed to evolve, and taking into account that
\be
\la z_{1}|z_{2}\ra=\la w_{1}|w_{2}\ra=0\,,\quad
\la z_{1}|z_{2}]=-e^{-i\theta}\la z|z \ra\,,\quad
\la w_{1}|w_{2}]=-e^{-i\vphi}\la w|w \ra\,,
\ee
the equations of motion for $z_{1},w_{1}$ read:
\beq
\rd_{t}|z\ra &=&
-i(\lambda+2\bar{\gamma}e^{-i(\vphi+\theta)}){  2}Y\,|z\ra\,,
\\
\rd_{t}|w\ra &=&	
-i(\lambda+2\bar{\gamma}e^{-i(\vphi+\theta)}){  2}X\,|w\ra\,,
\eeq
while the equations of motion for $z_{2},w_{2}$ read:
\beq
\rd_{t}(e^{i\theta}|z]) &=&
-i(\lambda+2\bar{\gamma}e^{-i(\vphi+\theta)}){  2}Y\,e^{i\theta}|z]\,,
\\
\rd_{t}(e^{i\vphi}|w]) &=&	
-i(\lambda+2\bar{\gamma}e^{-i(\vphi+\theta)}){  2}X\,e^{i\vphi}|w]\,,
\eeq
from which we easily extract the evolution equations for the relative phases between the first and second link:
\beq
\rd_{t}\theta&=&-{  4}Y(\lambda+\gamma e^{i(\theta+\vphi)}+\bar{\gamma} e^{-i(\theta+\vphi)})\,, \\
\rd_{t}\vphi&=&-{  4}X(\lambda+\gamma e^{i(\theta+\vphi)}+\bar{\gamma} e^{-i(\theta+\vphi)})\,.
\eeq
We also compute the evolution equation for the vector norms,
\beq
\rd_{t}X&=&{  4}i(\gamma e^{i(\theta+\vphi)}-\bar{\gamma} e^{-i(\theta+\vphi)})XY\,, \\
\rd_{t}Y&=&{  4}i(\gamma e^{i(\theta+\vphi)}-\bar{\gamma} e^{-i(\theta+\vphi)})XY\,.
\eeq
Since $\rd_{t}(Y-X)=0$, this confirms that we can safely assume the norm matching condition $Y=X$ on the initial configurations and that this condition is then preserved during the evolution. In that case, we also have that the phase difference does not evolve, $\rd_{t}(\theta-\vphi)=0$, and we can focus on the dynamics of the total phase $\Theta\equiv (\theta+\vphi)$. For closed and matching configurations, we are thus left with a pair of equations of motion:
\beq
\rd_{t}\Theta&=&-{  8}(\lambda+\gamma e^{i\Theta}+\bar{\gamma} e^{-i\Theta}) X\,,
\label{evolTheta}\\
\rd_{t}X&=&{  4}i(\gamma e^{i\Theta}-\bar{\gamma} e^{-i\Theta})X^{2}\,.\label{evolX}
\eeq
We can proceed to a final simplification, if we assume that the Hamiltonian is actually a Hamiltonian constraint, i.e. if we focus on a trajectory with vanishing energy, $H=0$. Then, this imposes:
\be
H=0={  8}(\lambda+\gamma e^{i\Theta}+\bar{\gamma} e^{-i\Theta}) X^{2}\,.
\ee
In that case, the total phase $\Theta$ is constant and we are left with a single non-trivial equation:
\be
\f{\rd_{t}X}{X^{2}}={  4}i(\gamma e^{i\Theta}-\bar{\gamma} e^{-i\Theta})
\quad\Rightarrow\quad
X(t)=\f{-1}{{  4i}(\gamma e^{i\Theta}-\bar{\gamma} e^{-i\Theta})(t-t_{0})}=\f{1}{  8t \,\text{Im}(\gamma e^{i\Theta})+1/X_0}
\,.\label{eq:H=0Ez}
\ee
Since the phases do not evolve, we can assume that they vanish, $\theta=\vphi=\Theta=0$, in which case the spinor simply evolves as $|z\ra = \sqrt{X(t)/X_0}\,|z_{0}\ra$ for any fixed initial value $|z_{0}\ra$. Notice also that the evolution of $X(t)$ remains constant if $\text{Im}(\gamma e^{i\Theta})=0$.

If the energy does not vanish, the equations of motion can be written as decoupled 2nd order differential equations if we take into account that the energy itself is still a constant of motion, $\rd_{t}H=0$:
\be
\rd_{t}^{2}(X^{-1})=\f{  4H}{X}\left(\lambda- \f{H}{8X^2}\right)
\,,\qquad
\rd_{t}^{2}{\Theta}={  4}iH(\gamma e^{i\Theta}-\bar{\gamma} e^{-i\Theta})
\,.
\label{eq:2edgesHnotzero}
\ee
In order to explore the dependence of the solutions of equations \eqref{eq:2edgesHnotzero} with the initial values and coupling constants we have studied a large number ($\sim 10^4$) of numerical solutions for different values of the parameters. We plotted the results (see figure \ref{fig_3regimes} for an example) and looked for different patterns in them (see appendix \ref{appNumplots} for more details about the procedure). We found the following interesting results:
\begin{description}
    \item[Result 1] If $\lambda^2>4|\gamma|^2$, the evolution oscillates.
    \item[Result 2] If $\lambda^2\leq 4|\gamma|^2$, the evolution diverges.
     \item[Result 3] The behavior of the solutions does not depend on the initial conditions on the spinors. However, such initial values will affect the oscillation frequency and the value of the time at which the divergent regimes diverge.
    \item[Comment] In some specific cases the evolution is constant (a behavior similar to the case of the solutions of equation \ref{eq:H=0Ez} when $\gamma e^{i\Theta}\in\mathbb{R}$, where the divergent regime became constant). 
\end{description}

\begin{figure}[htb]
\begin{center}
\includegraphics[width=0.8\textwidth]{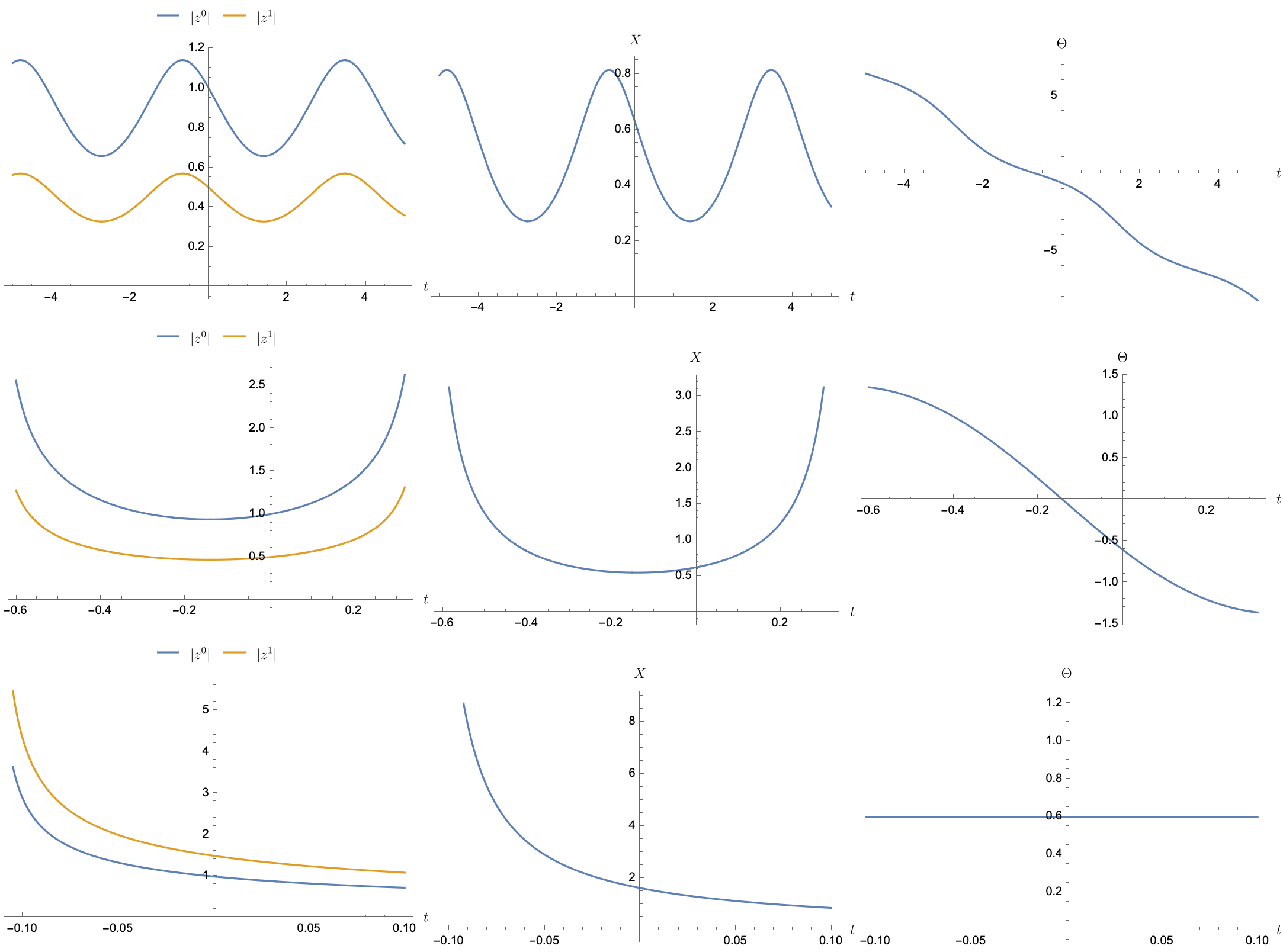}
\caption{The evolution of the spinorial variables given by equations \eqref{eq:2edgesHnotzero} are represented for the different regimes. In each column it is shown the evolution of the spinors of the vertex $\alpha$, the total area $X$ and the evolution of the total phase $\Theta$. On the other hand, in the first row an example of the oscillatory regime (with $\lambda^2>4|\gamma|^2$) is represented, in the second row we illustrate a case in the divergent regime, given by $\lambda^2\leq4|\gamma|^2$, and in the third one the evolution for $H=0$, which is also divergent in the spinor components and area.}
\label{fig_3regimes}
\end{center}
\end{figure}

\subsection{Lattice Gauge Theory Dynamics}

The lattice gauge theory Hamiltonian \eqref{eq:LGTHam} takes the following form for the case with 2 edges that we are studying:
\be
\tH ={  4}\left(
2\lambda
+\lambda\f{\la z_{1}|z_{2}\ra\la w_{1}|w_{2}\ra+\la z_{2}|z_{1}\ra\la w_{2}|w_{1}\ra}
{\sqrt{\la z_{1}|z_{1}\ra\la w_{1}|w_{1}\ra\la z_{2}|z_{2}\ra\la w_{2}|w_{2}\ra}}
+2\f{\gamma[ z_{1}|z_{2}\ra[ w_{1}|w_{2}\ra+\bar{\gamma}\la z_{1}|z_{2}]\la w_{1}|w_{2}]}
{\sqrt{\la z_{1}|z_{1}\ra\la w_{1}|w_{1}\ra\la z_{2}|z_{2}\ra\la w_{2}|w_{2}\ra}}\right)
\,.
\ee
Therefore, the equations of motion for the spinors read:
\be
\rd_{t}|z_{1}\ra
=
-\{\tH,|z_{1}\ra\}
=
-{  4i}\f{\lambda\la w_{1}|w_{2}\ra\,|z_{2}\ra+2\bar{\gamma}\la w_{1}|w_{2}]\,|z_{2}]}
{\sqrt{\la z_{1}|z_{1}\ra\la w_{1}|w_{1}\ra\la z_{2}|z_{2}\ra\la w_{2}|w_{2}\ra}}
+\f i2(\tH-{  8}\lambda)\f{|z_{1}\ra}{\la z_{1}|z_{1}\ra}
\,,
\ee
and similarly for the other spinors. For closed and matching configurations (which are still preserved under evolution), this simplifies to:
\be
\rd_{t}|z\ra
=
\f iX\left(
\f{1}{{  4}}\tH-{  2}\lambda-{  4}\bar{\gamma}e^{ -i\Theta}
\right)\,|z\ra\,{ = \f{2i}{X}\left(\gamma e^{i\Theta}-\bar{\gamma} e^{-i\Theta}\right)\,|z\ra,}
\ee
leading to the norm evolution equations:
\be
\rd_{t}X={  4}i(\gamma e^{i\Theta}-\bar{\gamma} e^{-i\Theta})
\,,
\ee
without any power of the norm on the right-hand side.

The equations of motion for $|z_{2}\ra$ read:
\be
\rd_{t}(e^{i\theta}|z]) =
\f{2i}{X}\left(\gamma e^{i\Theta}-\bar{\gamma} e^{-i\Theta}\right)e^{i\theta}|z],
\ee
and similarly for $\omega_2$. From these equations, we easily extract the evolution equations for the relative phases between the first and second link:
\be
\rd_{t}\theta=0=
\rd_{t}\vphi=\rd_{t}\Theta,
\ee
so the phases remain constant under evolution (for any value of the energy) and the norm $X$ evolves linearly with $t$. Therefore, the evolution with this Hamiltonian does not show the oscillatory and divergent regimes that we obtained for the LQG Hamiltonian \eqref{eq:LQGham}.

\subsection{Intermediate Hamiltonian Dynamics}

The intermediate Hamiltonian introduced in equation \eqref{IntermediateHam} takes the following form in the case of $N=2$ edges:
\begin{equation}
\begin{alignedat}{2}
H_\text{inter} = \frac{2}{\sqrt{\left(\la z_1\lvert z_1\ra+ \la z_2\lvert z_2\ra\right)\left(\la \omega_1\lvert \omega_1\ra+\la \omega_2\lvert \omega_2\ra\right)}}\bigg[&\lambda\big(\la z_1\lvert z_1\ra \la\omega_1\lvert\omega_1\ra + \la z_2\lvert z_2\ra \la\omega_2\lvert\omega_2\ra \big)\\
&+\gamma\big([z_1\lvert z_2\ra[\omega_1\lvert\omega_2\ra+[z_2\lvert z_1\ra[\omega_2\lvert\omega_1\ra\big)\\
&+\bar{\gamma}\big(\la z_2\lvert z_1]\la\omega_2\lvert\omega_1]+\la z_1\lvert z_2]\la\omega_1\lvert\omega_2]\big)\bigg]\,,
\end{alignedat}
\end{equation}
and the equations of motion for the variables at the vertex $\alpha$: 
\begin{eqnarray}
   && \rd_t\lvert z\ra = \f{i}{2}\sqrt{\frac{Y}{X}}\left(-\lambda+\gamma e^{i\Theta}-3\bar{\gamma}e^{-i\Theta}\right)\lvert z\ra,\\
   &&  \rd_t \theta = -\f{H_\text{inter}}{4X},\\
   && \rd_t X = 2 i \sqrt{XY}\left(\gamma e^{i\Theta}-\bar{\gamma}e^{-i\Theta}\right), 
\end{eqnarray}
and similarly for the vertex $\beta$.

Therefore, after applying the matching constraint we end up with the following system of differential equations:
\beq
\rd_t X &=& -4\,\text{Im}\!\left(\gamma e^{i\Theta}\right) X,\label{eq:interm_eoms1}\\
\rd_t \Theta &=& -\f{H_\text{inter}}{2X}\,.
\label{eq:interm_eoms2}
\eeq 
In order to study the emergence from equations \eqref{eq:interm_eoms1} and \eqref{eq:interm_eoms2}  of possible regimes (oscillatory or divergent), we have repeated the procedure explained in Appendix \ref{appNumplots} for this case. The results that we obtained are exactly the same as the results for LQG Hamiltonian \eqref{eq:LQGHam_2edges}, for which the regimes only depend on  $\text{Sign}\left(\lambda^2 - 4|\gamma|^2\right)$.

\section{LQG evolution for the isosceles tetrahedron}\label{sec:tetrahedron}

We now study the evolution given by the LQG Hamiltonian (\ref{eq:LQGham}) for  configurations  representing two matching isosceles tetrahedra. This model  has in general non-zero volume. This apparently simple case is important given that the tetrahedra representing the geometry dual to the vertices represents the simplest possible chunk of non-zero volume within the LQG theory and that we can explore fluctuations of volume and shape, i.e. expansion and shear dynamics of the geometry.

\begin{figure}[htb]
\begin{center}
\includegraphics[width=0.4\textwidth]{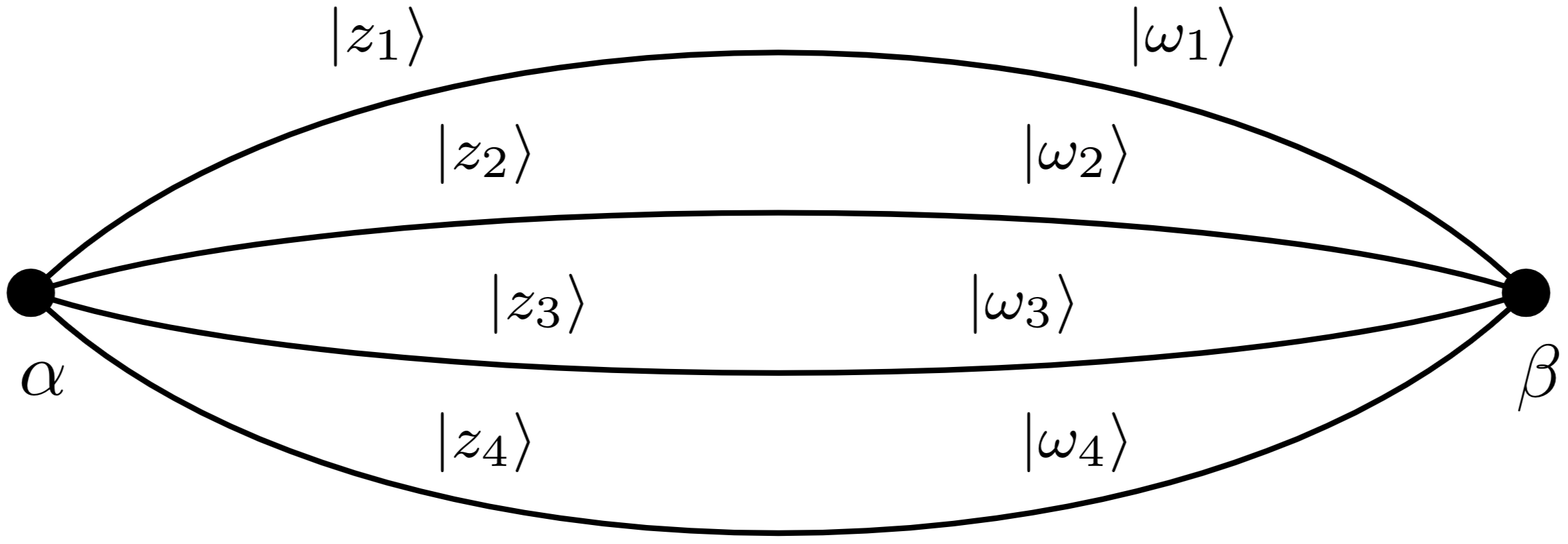}
\caption{The 2-vertex model with $N=4$ edges. Provided the parametrization of the spinors given by \eqref{eq:isoscelesspinors}, this graph corresponds to two isoceles tetrahedra attached by their faces satisfying the matching constraint (equal areas of the matching faces).}
\label{fig:4edges}
\end{center}
\end{figure}

Let us indeed consider the 2-vertex model with $N=4$ edges (see figure \ref{fig:4edges}) with the following parametrization of the spinors:

\begin{equation}
    \begin{aligned}
    &|z_{1}\rangle =\begin{pmatrix} \alpha\\\beta \end{pmatrix},
    &|z_{3}\rangle &=\begin{pmatrix} \xi\\\delta e^{i\theta} \end{pmatrix}, &|\omega_{1}\rangle &=\begin{pmatrix} \alpha\\\beta \end{pmatrix}, &|\omega_{3}\rangle &=\begin{pmatrix} \xi\\\delta e^{i\varphi} \end{pmatrix},\\
    &|z_{2}\rangle =\begin{pmatrix} \alpha\\-\beta \end{pmatrix}, 
    &|z_{4}\rangle &=\begin{pmatrix} \xi\\-\delta e^{i\theta}\\
    \end{pmatrix},\qquad\quad
    &|\omega_{2}\rangle & =\begin{pmatrix} \alpha\\-\beta \end{pmatrix}, &|\omega_{4}\rangle &=\begin{pmatrix} \xi\\-\delta e^{i\varphi} \end{pmatrix},\\
\end{aligned}
\label{eq:isoscelesspinors}
\end{equation}
where $\theta$ and $\varphi$ are constants. Imposing the closure constraint, we can determine the value of one of the parameters, say $\delta$:
\begin{equation}
    |\delta|=\sqrt{|\alpha|^2+|\xi|^2-|\beta|^2}.\label{eq:delta0}
\end{equation}

\subsection{Evolution equations for the spinors}

Using the description for the isosceles tetrahedron given by equation \eqref{eq:isoscelesspinors} and following the equations of motion \eqref{eomLQGham} corresponding to the LQG Hamiltonian \eqref{eq:LQGham}, the following evolution equations for the parameters are obtained:
\begin{align}
    &\dot{\alpha}=-2i\lambda\bar{\alpha}\left(\alpha^2+\xi^2\right)-4i\bar{\gamma}\bar{\alpha}\left(\bar{\beta}^2+\bar{\delta}^2e^{-i(\theta+\varphi)}\right),\\
    &\dot{\beta}=-2i\lambda\bar{\beta}\left(\beta^2+\delta^2 e^{i(\theta+\varphi)}\right)-4i\bar{\gamma}\bar{\beta}\left(\bar{\alpha}^2+\bar{\xi}^2\right),\\
    &\dot{\xi}=-2i\lambda\bar{\xi}\left(\alpha^2+\xi^2\right)-4i\bar{\gamma}\bar{\xi}\left(\bar{\beta}^2+\bar{\delta}^2e^{-i(\theta+\varphi)}\right),\\
    &\dot{\delta}=-2i\lambda\bar{\delta}e^{-i(\varphi+\theta)}\left(\beta^2+\delta^2 e^{i(\theta+\varphi)}\right)-4i\bar{\gamma}\bar{\delta}e^{-i(\varphi+\theta)}\left(\bar{\alpha}^2+\bar{\xi}^2\right).
\end{align}
Notice that the equations for this model do not allow for the creation of edges; if one variable is zero at the beginning, it will remain zero.

\begin{figure}[htb]
\begin{center}
\includegraphics[width=0.8\textwidth]{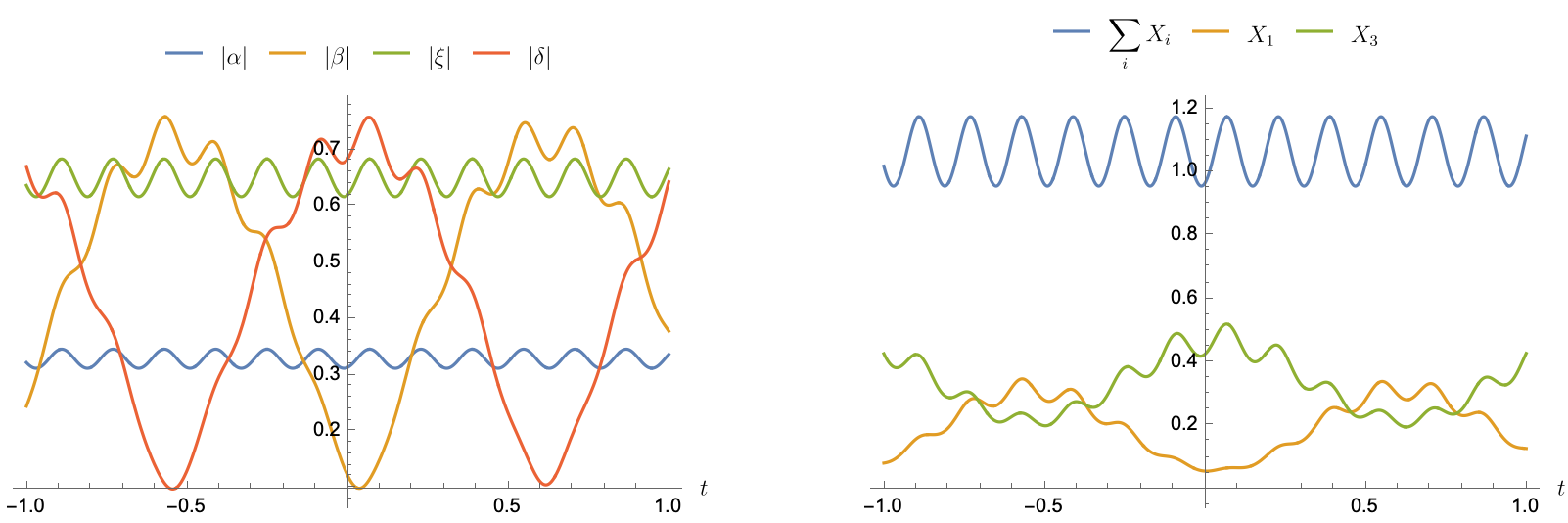}
\caption{The evolution of the spinor components (left) and the areas (right) in the oscillatory regime. In the case of the areas, only the total area and the areas of two of the faces are represented (the other two areas are equal to the ones plotted because of the symmetry of the isosceles tetrahedra).}
\label{fig_tetra_osc}
\end{center}
\end{figure}

In order to study the different solutions of the evolution equations for this case, we have also proceeded with a systematic strategy by calculating numerous plots for the evolution of the total area for different initial values and coupling constants, as done in the bivalent 2-vertex model (section \ref{sec:2vertexmodel}). Remarkably, the same regimes as in the previous case (also found for the symmetry reduced model of the reference \cite{Borja:2010rc}) are obtained. Therefore, the same results and comments are in order here: if $\lambda^2>4|\gamma|^2$ the evolution oscillates, for $\lambda^2\leq 4|\gamma|^2$ the evolution diverges, there is no dependence of the behavior with the initial values of the spinors and for certain specific cases a constant evolution is found. As an example, the behavior of the isosceles tetrahedra in the oscillatory regime is plotted in figure \ref{fig_tetra_osc}.

\subsection{Quadrupole moment}

The quadrupole moment associated with a polyhedron gives us direct information about its shape and may provide a first approximation for the volume. We define the quadrupole moment as \cite{Goeller:2018jaj}:
\be
T^{ab}=\sum_{i=1}^4\f{1}{X_i}X^a_iX^b_i=\f12\sum_{i=1}^4\f{1}{\la z_i|z_i\ra}\la z_i|\sigma^a|z_i\ra\la z_i|\sigma^b|z_i\ra,\label{eq:quadrupole}
\ee
and its trace\footnote{It is also possible to define a traceless quadrupole moment \cite{Goeller:2018jaj}.} is the total area around the node: $\tr\, T=\sum_i^4X_i$.

\begin{figure}[htb]
    \centering
    \includegraphics[width=0.9\textwidth]{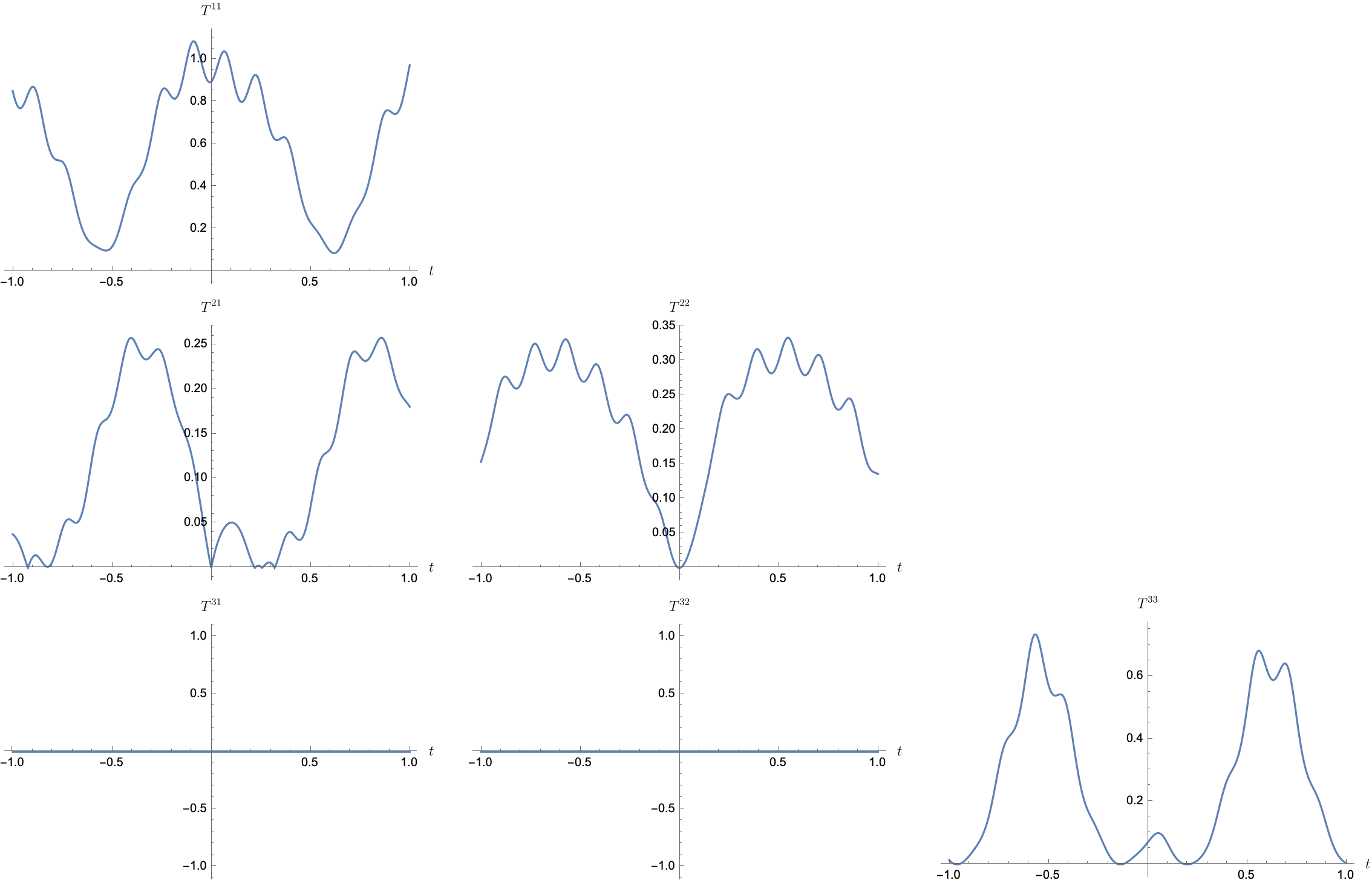}
    \caption{The evolution of the different components of the quadrupole are plotted for the isosceles tetrahedron (given that it is a symmetric matrix, we only plot the diagonal and the elements under the diagonal). Notice that, in this case, the elements $T^{31}=0=T^{32}$ as a consequence of the parametrization \eqref{eq:isoscelesspinors} of the isosceles tetrahedron.}
    \label{fig:tetra/3}
\end{figure}

\begin{figure}[htb]
  \includegraphics[width=0.4\textwidth]{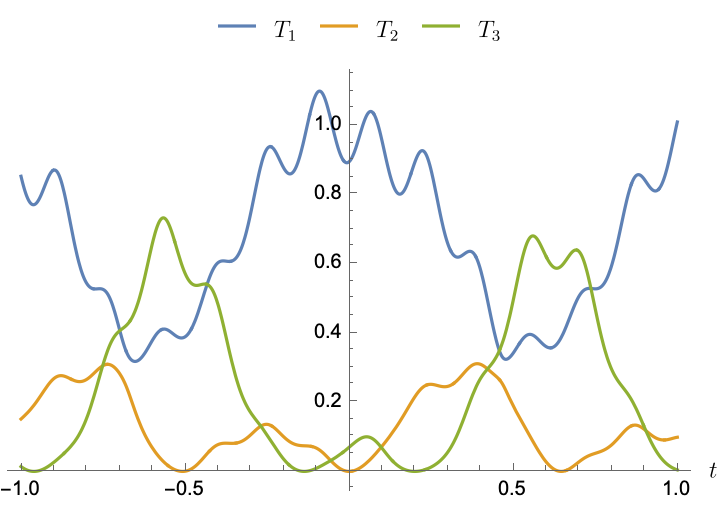}
  \caption{Evolution of the eigenvalues of the quadrupole moment $T^{ab}$, given by equation \eqref{eq:quadrupole}, for the oscillatory regime considered also in figure \ref{fig:tetra/3}.}
  \label{fig:tetra/4}
\end{figure}

\begin{figure}[htb]
  \includegraphics[width=0.4\textwidth]{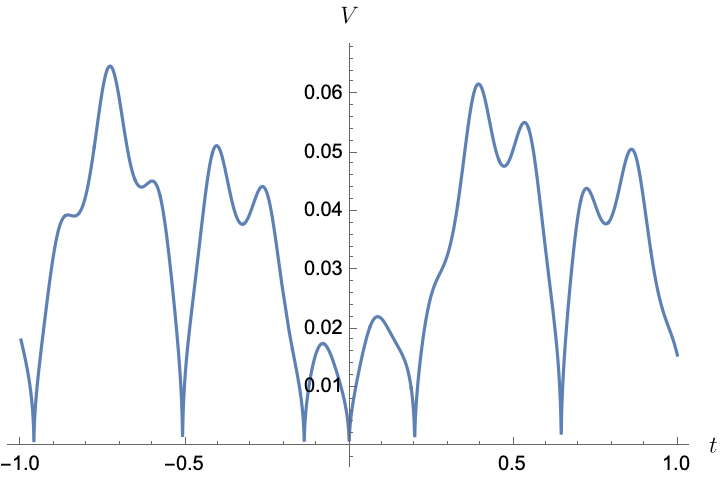}
  \caption{Evolution of the volume of the isosceles tetrahedron (given by equation \ref{eq:volume_tetra}) in the same oscillatory regime considered in figures \ref{fig:tetra/3} and \ref{fig:tetra/4}.}
  \label{fig:tetra/5}
\end{figure}

Given that the quadrupole may be written in terms of the spinors, we may compute the evolution of the quadrupole moment and obtain information about its eigenvalues and principal axes. For the case of the isosceles tetrahedron, we obtain the evolution of the quadrupole moment given in figure \ref{fig:tetra/3}, where, for the choice of coupling constants taken, an oscillatory behavior may be recognized. 
Furthermore, looking at the evolution of its eigenvalues (figure \ref{fig:tetra/4}), we observe that the distribution of the area will evolve irregularly along the three main directions. In fact, for the tetrahedron it is easy to compute its volume $V$, that is given by \cite{Goeller:2018jaj}:
\begin{align}
V^{2}=\frac{2}{9}\,|\vec{X}_1\cdot(\vec{X}_2\wedge\vec{X}_3)|.
\label{eq:volume_tetra}
\end{align}
As the volume is given in terms of the normal vectors (that may be written in terms of the spinors) we can study the evolution of the volume exactly. In figure \ref{fig:tetra/5} the evolution of the volume is represented and, as expected from the results for the quadrupole moment, the volume oscillates. The time values where zero volume is found correspond to the geometrical situations where two normal vectors are parallel. As we will observe in section \ref{sec:four}, for the case of polyhedra with more faces (graphs with larger number of links) it will be more difficult to find a combination of normal vectors for which the resulting polyhedron has zero volume.

\subsection{Reconstruction of the tetrahedron}
\label{sec:reconstruction_tetra}

\begin{figure}[htb]
\begin{center}
\includegraphics[width=0.35\textwidth]{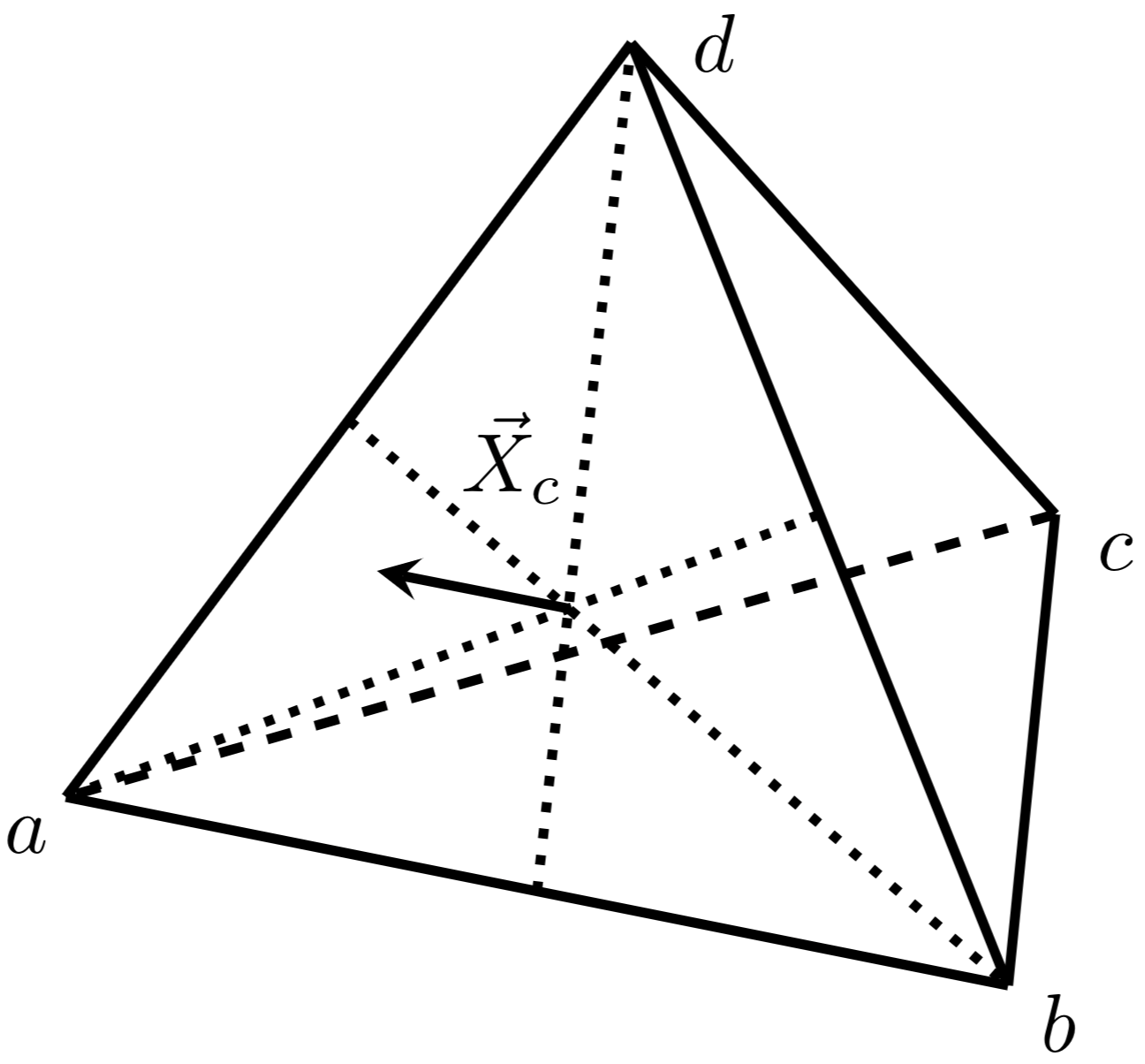}
\caption{An example of a tetrahedron. The normal vector to the face opposed to $c$ will be $\vec{X}_c$ and its modulus will be the area of the face $\widetriangle{abd}$.}
\label{fig:flattetra}
\end{center}
\end{figure}

We  review the reconstruction of the tetrahedron from the normal vectors $\vec{X}$ following \cite{Charles:2015lva}. First, we need to locate its vertices, which we  label by the normal vectors at the opposite face. To ease the notation (using the same notation as in \cite{Charles:2015lva}), we define $\vec{X}_a\equiv\vec{X}_1$, $\vec{X}_b\equiv\vec{X}_2$, $\vec{X}_c\equiv\vec{X}_3$ and $\vec{X}_d\equiv\vec{X}_4$. Naming the vertices $a$, $b$, $c$ and $d$, the normal vector $\vec{X}_c$  belongs to the face opposed to the vertex $c$, as shown in figure \ref{fig:flattetra}.

\begin{figure}[htb]
\begin{center}
\includegraphics[width=0.9\textwidth]{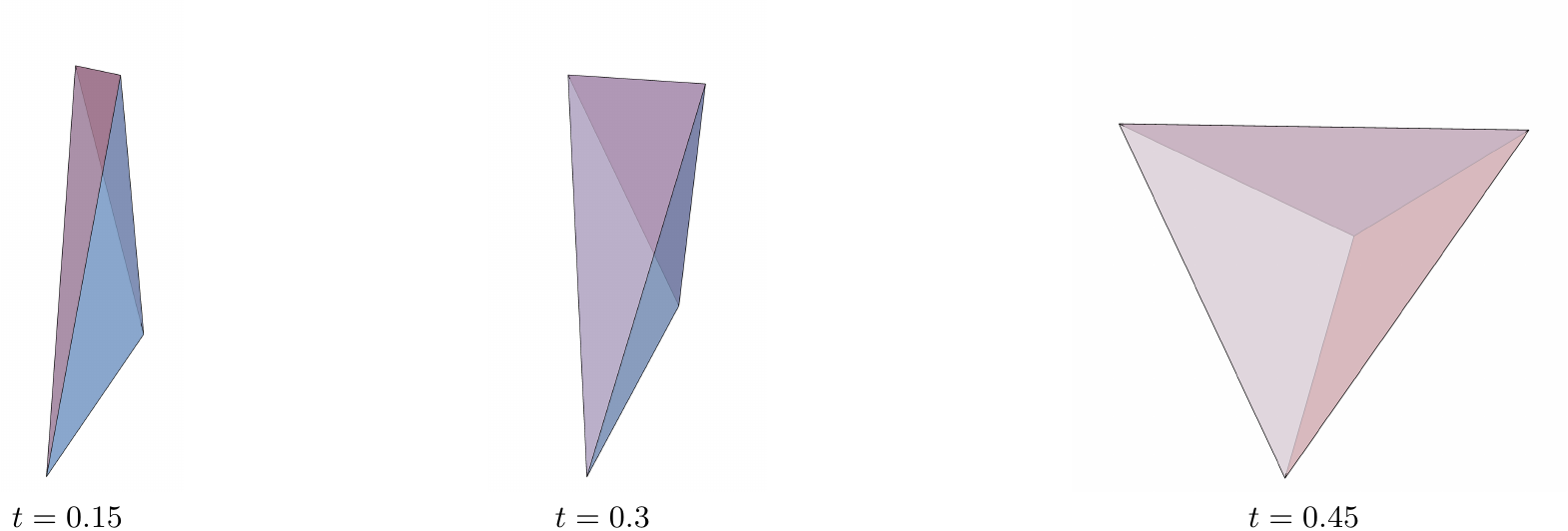}
\caption{Images of the reconstructed isosceles tetrahedron for three different time snaps corresponding to three different values of the volume in figure \ref{fig:tetra/5}.}
    \label{fig:tetra3d}
\end{center}
\end{figure}

Now, the relationship between the edges of the tetrahedron and its normals is given by \cite{Charles:2015lva}:
\begin{equation}
    \overrightarrow{ad}=\sqrt{\frac{2}{U}}\vec{X}_b\times\vec{X}_c,\quad\overrightarrow{ab}=\sqrt{\frac{2}{U}}\vec{X}_c\times\vec{X}_d,\quad\overrightarrow{ac}=\sqrt{\frac{2}{U}}\vec{X}_d\times\vec{X}_b,\label{eq:edgesoftetra}
\end{equation}
where $U=9V^2/2$. This  uniquely characterizes the tetrahedron. Therefore, if we fix the vertex $a$, the other vertices are fixed by the relations (\ref{eq:edgesoftetra}). In order to ease the graphical visualization, we  locate the vertex $a$ so that the centroid $G$ of the tetrahedron is fixed at the origin $O=(0,0,0)$ at any time. The centroid is given by:
\begin{align}
    G=\frac{a+b+c+d}{4}.
\end{align}
If we solve for $a$ and set $G=(0,0,0)$, we obtain the position of each vertex in terms of $\overrightarrow{ad}$, $\overrightarrow{ab}$ and $\overrightarrow{ac}$:
\begin{align}
    &d=\f{3}{4}\vec{ad}-\f{1}{4}\vec{ab}-\f{1}{4}\vec{ac},\label{eq:reconstruction_analytic1}\\
    &c=\f{2}{3}\vec{ac}-\f{1}{3}\vec{ab}-\f{1}{3}d,\label{eq:reconstruction_analytic2}\\
    &b=\frac{1}{2}(-c-d+\overrightarrow{ab}),\label{eq:reconstruction_analytic3}\\
    &a=-b-c-d.\label{eq:reconstruction_analytic4}
\end{align}
Once we have set the position of the vertices, we can illustrate our model with 3d images. As we can see in figure \ref{fig:tetra3d}, the tetrahedron  rotates and its faces  change their area and shape and, thus, the volume  also evolves (as already seen in figure \ref{fig:tetra/5}).

\section{LQG Hamiltonian dynamics for models with arbitrary number of edges}
\label{sec:four}

Up to now we have considered specific models parametrized in a simple way in order to ease the study of the equations of motion. In this section, we  consider completely general parametrizations of the spinors for 2-vertex models with arbitrary number of edges.

One of the main difficulties in dealing with arbitrary configurations for the spinors is that the closure constraint must be satisfied. We now present a method, based on the results of \cite{Livine:2012cv,Livine:2013tsa}, to generate closed arbitrary configurations of the spinors as initial values for the differential equations of motion. Then, we  solve numerically such equations, reconstruct the polyhedron for any moment of the evolution and study the evolution of its total area and volume.

Finally, we  consider a more general Hamiltonian with different coupling constants for different pairs of edges of the graph.

\subsection{Generating and evolving a random configuration on the 2-vertex graph}\label{subsec:generalrandom}

In order to obtain arbitrary configurations of spinors satisfying the closure constraint, we  use the toolbox of operations on collections of spinors detailed in appendix \ref{app:toolbox}. Then, we  run numerical simulations of the LQG Hamiltonian for the 2-vertex model on this configuration in order to solve the equations of motion and to obtain the evolution of the polyhedra (using the reconstruction algorithm discussed in \cite{Sellaroli:2017wwc,Bianchi:2010gc}) and of its volume.

The procedure is the following:
\begin{enumerate}
\item We  choose $N$ random spinors. First, we  consider random vectors on the unit sphere multiplied by weights that are chosen following a Gaussian distribution. Then, we  obtain the corresponding spinors associated with each vector \cite{Borja:2010rc}.
\item We  boost the spinors to their corresponding closed configuration (see appendix \ref{app:toolbox}). Those are the spinors $|z_{i}\ra$ around the vertex $\alpha$.
\item In order to obtain random spinors for the vertex $\beta$, we  pseudo-randomly deform them while keeping the individual areas of the faces fixed using the scalar product Hamiltonian flows defined in appendix \ref{app:toolbox}. This gives the spinors $w_{i}$ at the vertex $\beta$, satisfying the area-matching condition on the 2-vertex graph.
\end{enumerate}

Regarding the reconstruction of the polyhedra, the Minkowski theorem \cite{Minkowski1897} ensures the existence and uniqueness of the 3d convex polyhedron associated with the set of closed normal vectors. Nevertheless, it does not provide a method to compute it. For the simple case of the tetrahedron we may use the method \cite{Charles:2015lva} described in section \ref{sec:reconstruction_tetra}, but for polyhedra with more faces the reconstruction algorithm is much more intricate \cite{Bianchi:2010gc,Sellaroli:2017wwc}. 

\begin{figure}[ht]
\begin{center}
\includegraphics[width=0.8\textwidth]{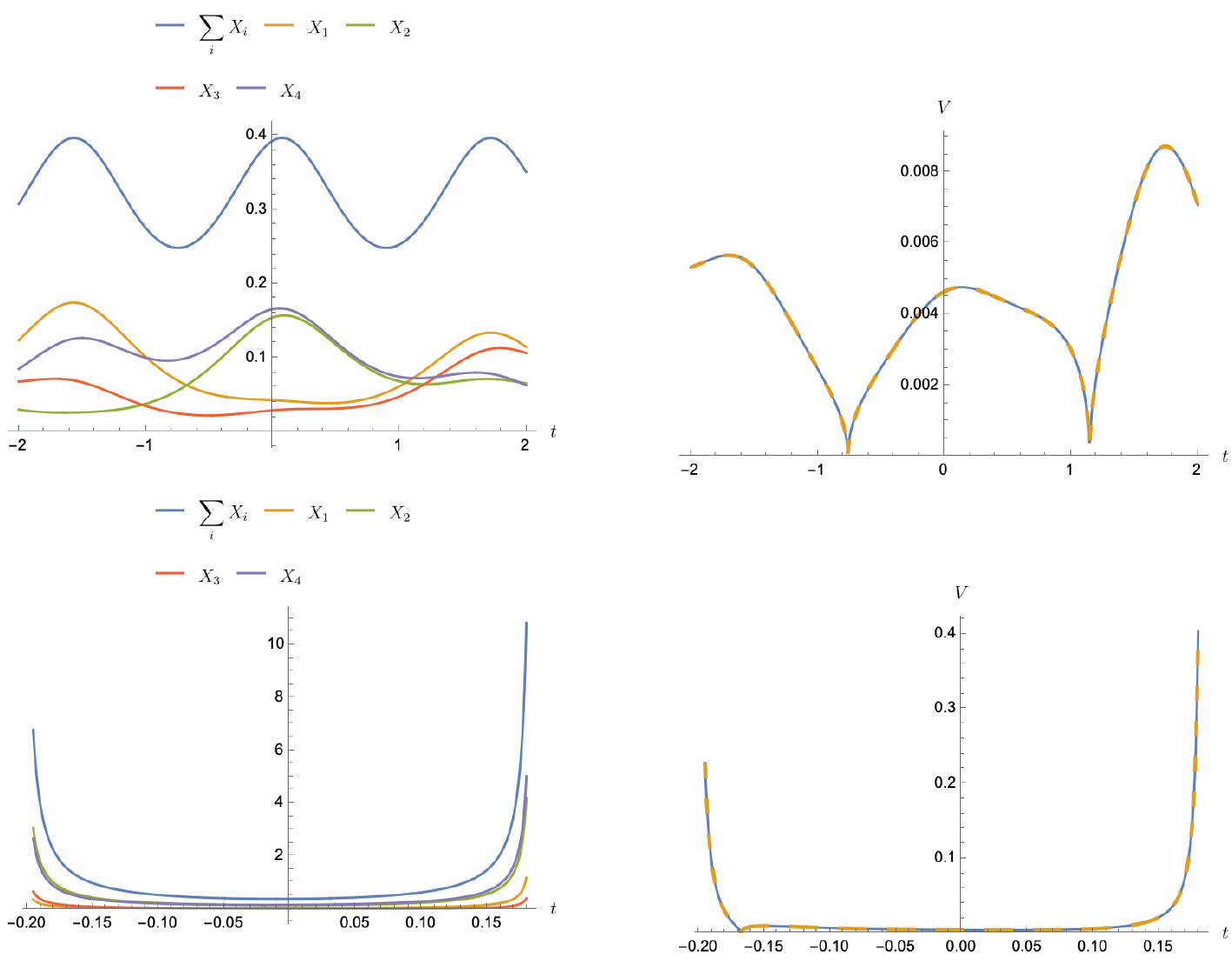}
\caption{Evolution of the areas (left) and volume (right) for a tetrahedron with randomly chosen initial values. The blue line of the volume corresponds to the result from the reconstruction algorithm in \cite{Sellaroli:2017wwc}, whereas the dashed orange line is the volume calculated using equation \eqref{eq:volume_tetra}. In the first row, the case for the oscillatory regime with $\lambda=5$ and $\gamma=2$ is represented, whereas in the second row we illustrate the divergent regime with $\lambda=2$ and $\gamma=5$.}
\label{fig:random_1}
\end{center}
\end{figure}

In our case, in order to compute the evolution of the volume, we have applied the algorithm presented by Sellaroli \cite{Sellaroli:2017wwc} and the associated implementation in \textit{Python}%
\footnote{Publicly available at \url{https://github.com/gsellaroli/polyhedrec}.}
(based on a method by Lasserre \cite{Lasserre1983AnAE}) at each step of the evolution, in order to compute numerically the volume and be able to plot its evolution. As a consistency check, for $N=4$ we have compared the reconstructed polyhedron as well as its volume using both Sellaroli's code and the analytical formulas from equations (\ref{eq:reconstruction_analytic1})-(\ref{eq:reconstruction_analytic4}) to check that they agree (figure \ref{fig:random_1}).

It is worth noting that, in this case, we recover again the same regimes we observed for the case with $N=2$ and for the isosceles tetrahedron, that is, the regimes also depend on $\text{Sign}(\lambda^2-4|\gamma|^2)$ and no dependence of the behavior with the initial values of the spinor has been found. In order to systematically check this result, we have proceeded in an analogous way as in the case with 2 edges or the isosceles tetrahedron. We have computed numerically the evolution of a large number of systems with different coupling constants and number of edges (see appendix \ref{appNumplots}) and the results that we have obtained reinforce those found for the simpler cases treated before.

\begin{figure}[htb]
\begin{center}
\includegraphics[width=0.4\textwidth]{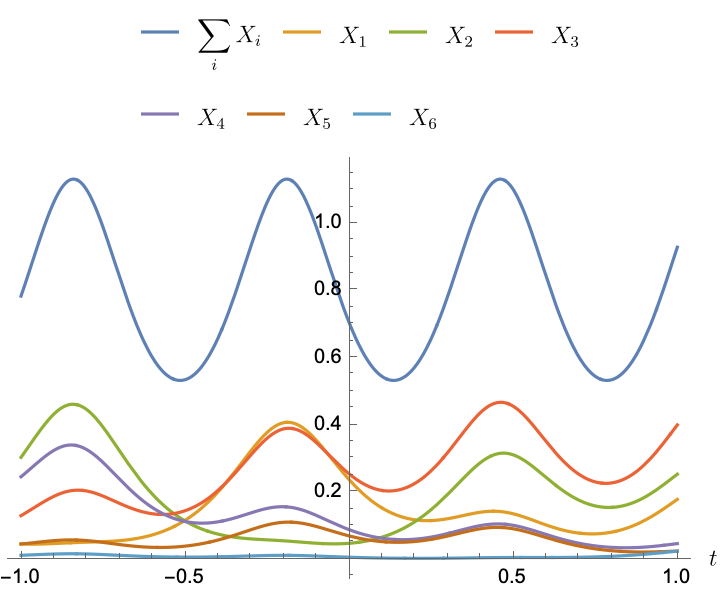}
\caption{Evolution of the areas in the oscillatory regime for a model with $N=6$ edges for a randomly chosen initial configuration of spinors.}
\label{fig:random_2}
\end{center}
\end{figure}

\begin{figure}[htb]
\begin{center}
\includegraphics[width=0.8\textwidth]{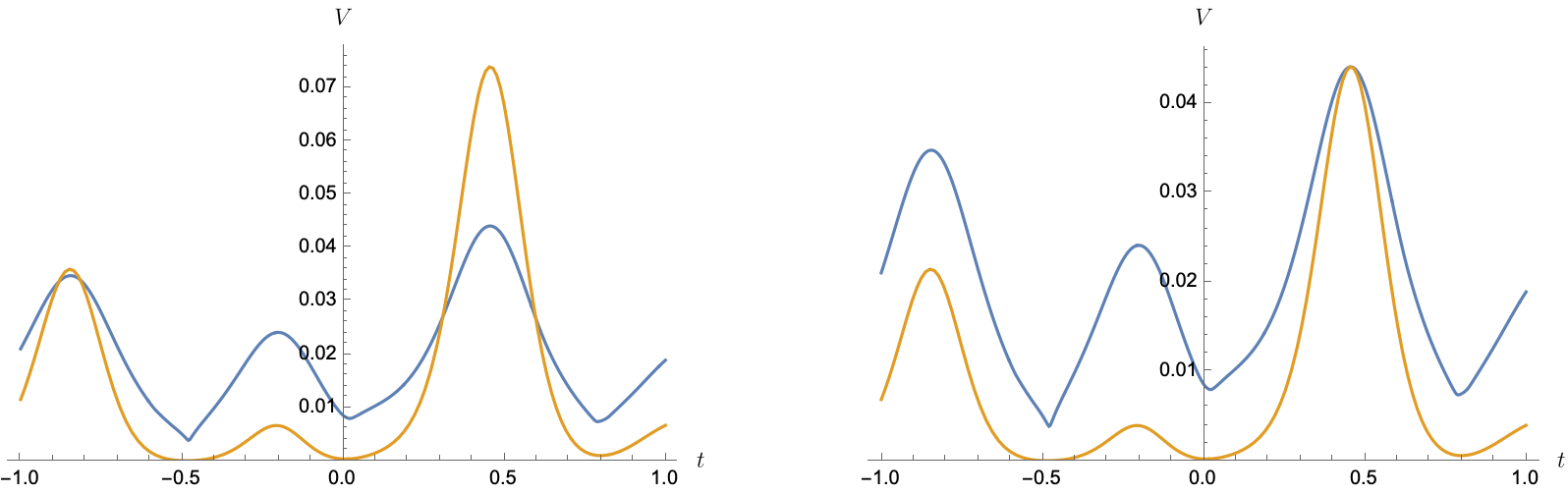}
\caption{We plot the evolution of the volume using the algorithm by Sellaroli (blue) and the approximate equation \eqref{eq:volumequadrupole} based on the eigenvalues of the quadrupole moment (orange). On the left-hand side the values obtained directly from the equation \eqref{eq:volumequadrupole} are used, whereas on the right-hand side the orange line is normalized to have the same maximum as the exact evolution of the volume (blue). This way, we can appreciate that they share the same tendencies.
}
\label{fig:random_4}
\end{center}
\end{figure}

Sellaroli's algorithm \cite{Sellaroli:2017wwc} is an excellent tool to find the vertices of a polyhedron and thus calculate the volume. Nevertheless, in order to ease the computational time when studying the evolution, specially for models with large number of edges (faces), we may approximate the volume with the eigenvalues of the quadrupole moment. Taking into account that the quadrupole moment has information about the distribution of area along the 3d object, the eigenvalues will provide a notion of which is the `amount of area' along every main direction (given by the corresponding eigenvector). Thus, we may use the formula for the volume of an ellipsoid to approximate the volume of our polyhedra, where the radii will be the three eigenvalues:
\begin{equation}
    V=\frac{4\pi}{3} T_{1}T_{2}T_{3},\label{eq:volumequadrupole}
\end{equation}
being $T_{1}$, $T_{2}$ and $T_{3}$ the eigenvalues of the quadrupole (given by equation \ref{eq:quadrupole}). As an example, we have studied the case of a random hexahedron in the oscillatory regime (see figure \ref{fig:random_2} for the evolution of the areas). For this case, we have computed the evolution of the volume in an exact way (using Sellaroli's method) and in an approximate way (using equation \ref{eq:volumequadrupole}). We can observe in figure \ref{fig:random_4} that, although the evolution based on equation \eqref{eq:volumequadrupole} does not provide an accurate numerical value for the volume, both plots have similar trends.

\begin{figure}[!htb]
\begin{center}
\includegraphics[width=0.9\textwidth]{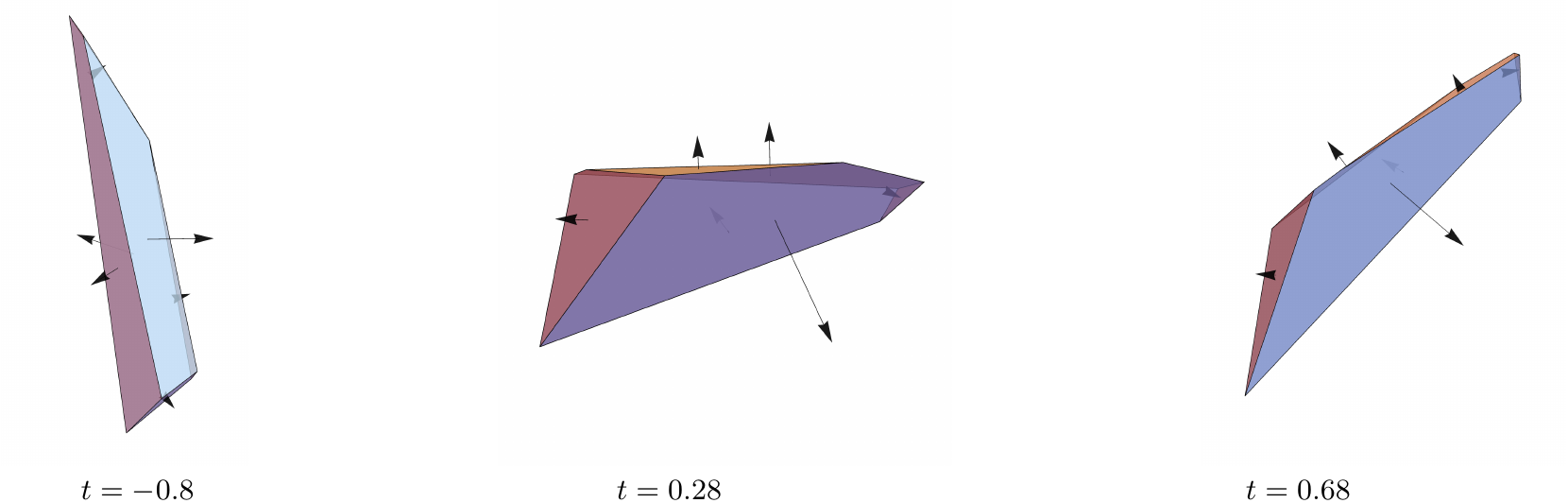}
 \caption{Images of the hexahedron corresponding to three different time snaps. The vectors arising from the faces are the normal vectors, whose norms correspond to the area of the corresponding faces.}
    \label{fig:recons/3}
\end{center}
\end{figure}

Finally, we may observe the actual evolution of the hexahedron given by the reconstruction algorithm at any time. In figure \ref{fig:recons/3} it is represented the hexahedron (with the normal vectors to the faces) at three different snaps of the evolution.

\begin{figure}[htb]
\begin{center}
\includegraphics[width=0.8\textwidth]{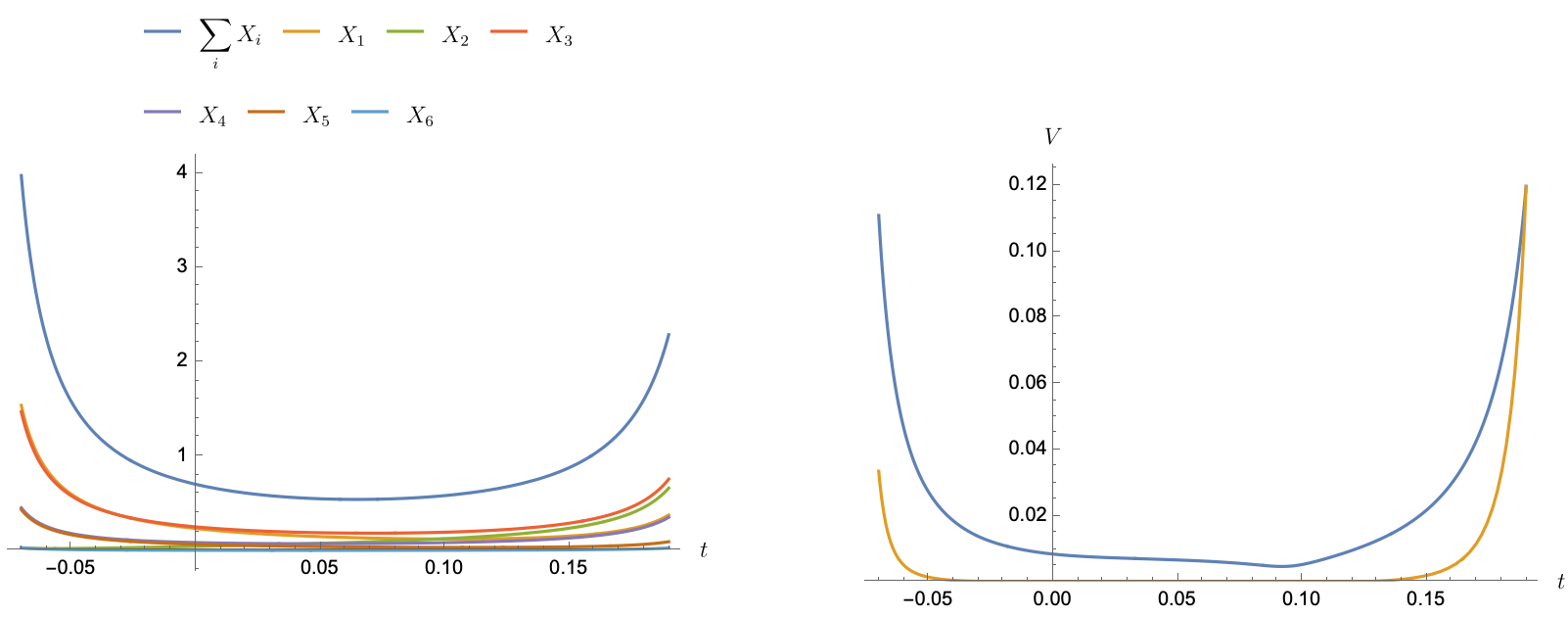}
\caption{On the left-hand side, the evolution of the areas for the divergent regime is represented, with the same initial values as in figure \ref{fig:random_2}, but with swapped coupling constants ($\lambda=2$ and $\gamma=5$). On the right-hand side, the evolution of the volume using the algorithm by Sellaroli (blue) and the approximate normalized volume (orange) using the eigenvalues of the quadrupole moment (equation \ref{eq:volumequadrupole}) are plotted.}
\label{fig:random_6}
\end{center}
\end{figure}

\begin{figure}[htb]
\begin{center}
\includegraphics[width=0.95\textwidth]{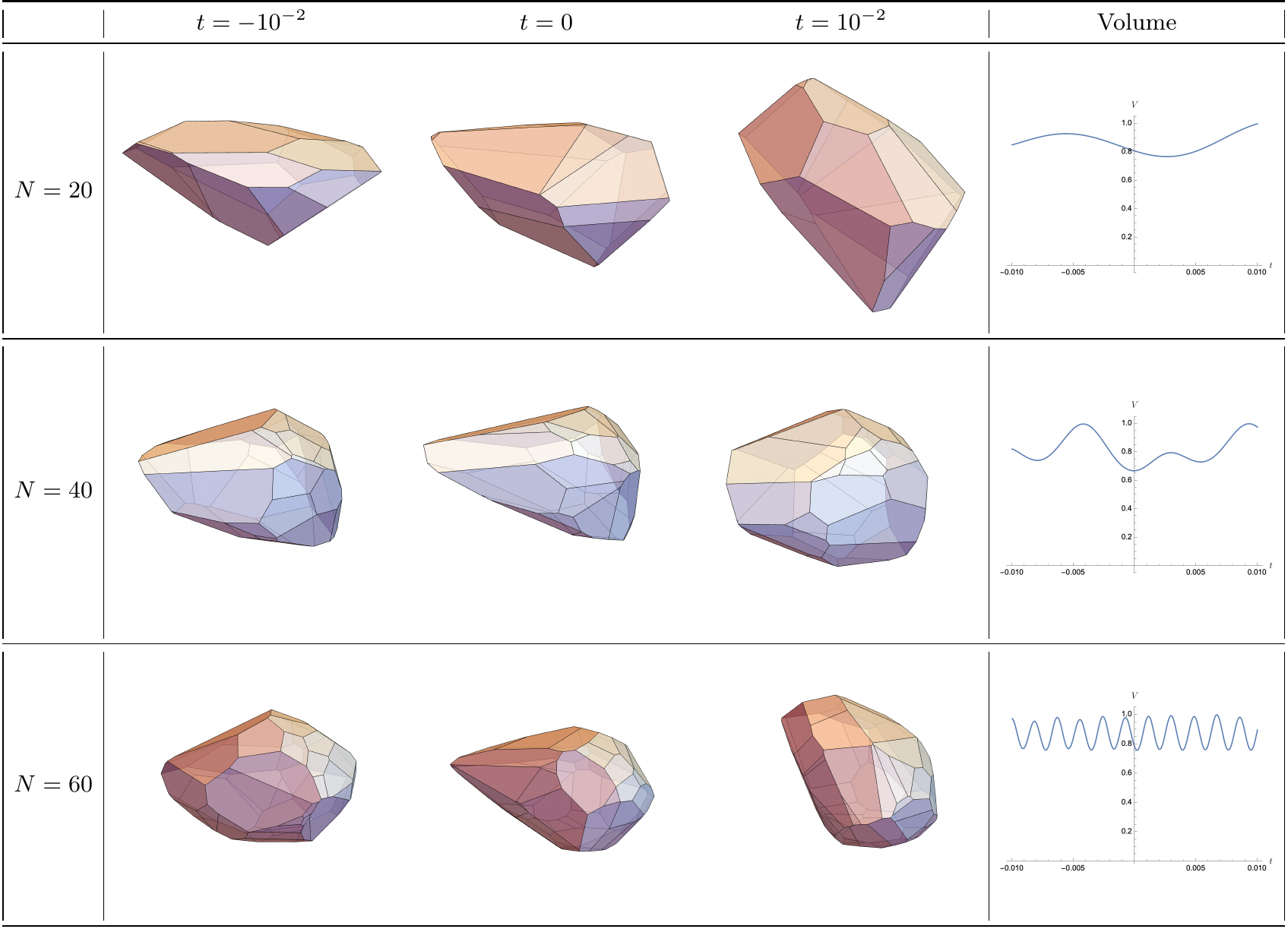}
 \caption{Table of polyhedra with different number of faces at three time snaps. The last column represents the evolution of the volume calculated using the approximation given by equation \eqref{eq:volumequadrupole}, normalized so that the maximum value is set to $V=1$. As previously explained, for large number of faces the volume is less likely to approach $V=0$.}
    \label{tbl:table_of_figures}
\end{center}
\end{figure}

Regarding the divergent regime, we may proceed analogously, and we find similar results (see figure \ref{fig:random_6}) for the hexahedron. Furthermore, proceeding in the same way, we may compute the evolution equations for any random polyhedra\footnote{
Although the reconstruction algorithm \cite{Sellaroli:2017wwc} is suitable for large number of faces, in order to study the evolution we have to solve numerically the equations of motion and, then, implement Sellaroli's algorithm at each step of the evolution. This process requires computational time. In our case, we have obtained results for polyhedra with a maximum of 60 faces. 
}
and, remarkably, we observe the emergence of the same regimes, depending only on $\text{Sign}(\lambda^2-4|\gamma|^2)$ (with no dependence on the initial values). For the cases with many faces, calculating the volume using Sellaroli's algorithm is a computationally expensive procedure, but we can take advantage of the approximation given by the eigenvalues of the quadrupole moment (equation \ref{eq:volumequadrupole}) in order to gain intuition about the behavior of the volume of a polyhedra with large number of faces (figure \ref{tbl:table_of_figures}).


\subsection{Edge-dependent coupling constants}
\label{sec:edge-dep}

As an extension of the results obtained for the LQG Hamiltonian (equation \ref{eq:LQGham}), we may also study the evolution generated by a more general Hamiltonian, considering coupling constants which depend on the edges:
\be
H_{\text{gen}}=\sum_{k,l}\lambda_{kl} E^{\alpha}_{kl}E^{\beta}_{kl}
+\gamma_{kl}  F^{\alpha}_{kl}F^{\beta}_{kl}
+\bar{\gamma}_{kl} \bF^{\alpha}_{kl}\bF^{\beta}_{kl}
\,.
\label{eq:Hamedge}
\ee
Notice that this Hamiltonian still commutes with the closure and matching constraints \cite{Borja:2010rc}. Also note that only the symmetric part of $\gamma_{kl}$ contributes to the dynamics and that $\lambda_{kl}=\lambda_{lk}$  in order for the Hamiltonian \eqref{eq:Hamedge} to be Hermitian.

\begin{figure}[htb]
\begin{center}
\includegraphics[width=0.8\textwidth]{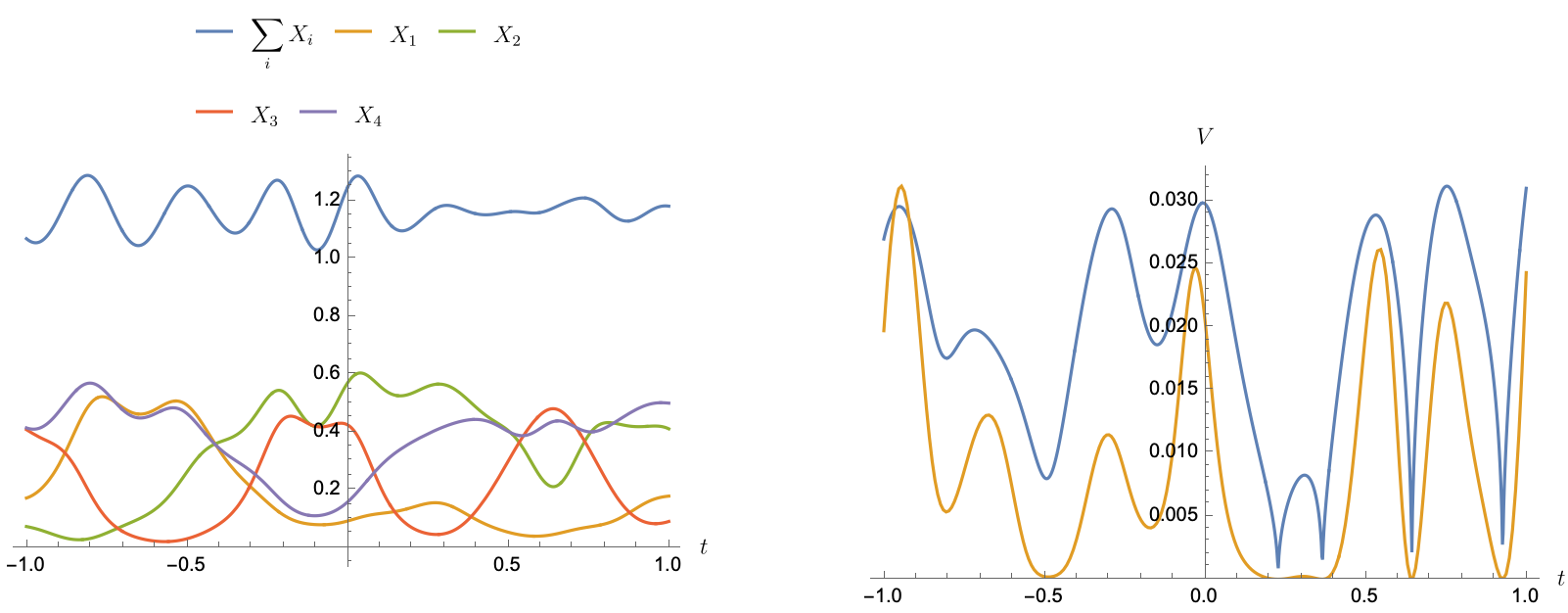}
\caption{On the left-hand side it is shown the evolution under the Hamiltonian \eqref{eq:Hamedge} of the individual areas of the faces for the random tetrahedron and its total area. On the right-hand side the volume is represented: the blue line corresponds to the reconstruction algorithm from \cite{Sellaroli:2017wwc}, whereas the orange line has been obtained from the quadrupole-eigenvalues approximation of equation \eqref{eq:volumequadrupole}. We may observe that, in the time interval considered, the evolution appears to oscillate, nevertheless a general description of the oscillatory behavior in terms of the values of the coupling constants has not been found.}
\label{fig:random_5}
\end{center}
\end{figure}

\begin{figure}[htb]
\begin{center}
\includegraphics[width=0.8\textwidth]{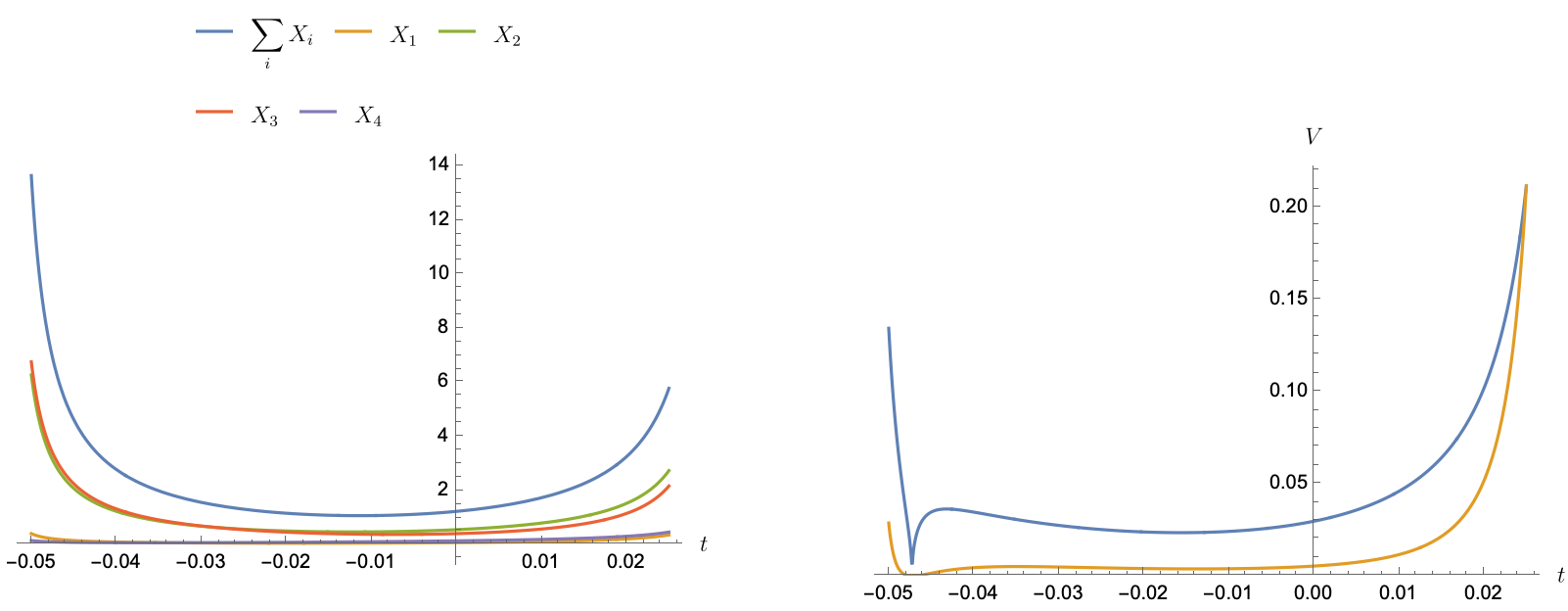}
\caption{Evolution of the areas (left) and the volume (right) in a divergent regime for the evolution generated by the Hamiltonian \eqref{eq:Hamedge}. The blue line corresponds to the reconstruction algorithm from \cite{Sellaroli:2017wwc}, whereas the orange line has been obtained from the quadrupole-eigenvalues approximation given by equation \eqref{eq:volumequadrupole}.}
\label{fig:randomdiv}
\end{center}
\end{figure}

We have studied the evolution generated by the Hamiltonian \eqref{eq:Hamedge} for different initial random configurations of the spinors, for different number of faces and for different values of the coupling constants. No recognisable regimes dependent on the coupling constants (as it was the case with the LQG Hamiltonian considered before) have been observed for the dynamics given by equation \eqref{eq:Hamedge}. Nevertheless, it seems that the oscillatory and divergent behaviors may appear for certain specific values of the parameters. As an example, in figure \ref{fig:random_5} we have plotted the evolution of a random tetrahedron with the following values of $\lambda_{kl}$ and $\gamma_{kl}$,
\be
\lambda_{kl}=\begin{pmatrix}
5&-6&7&-4\\
-6&5&-8&5\\
7&-8&8&-5\\
-4&5&-5&5
\end{pmatrix},\qquad
\gamma_{kl}=\begin{pmatrix}
0.3i&0.5i&0.2i&0.5\\
0.5i&0.2+0.5i&0.1+i&-0.1i\\
0.2i&0.1+i&2i&-0.5i\\
0.5&-0.1i&-0.5i&0.2i
\end{pmatrix}.
\ee
In this case, the evolution is oscillatory but clearly non-periodic. In fact, it seems that choosing edge-dependent coupling constants makes the fluctuations of area and volume more irregular (as may be intuitively expected). On the other hand, if we choose the following coupling constants,
\be
\lambda_{kl}=\gamma_{kl}=\begin{pmatrix}
5&-6&7&-4\\
-6&5&-8&5\\
7&-8&8&-5\\
-4&5&-5&5
\end{pmatrix},
\ee
we observe that the evolution diverges (see figure \ref{fig:randomdiv}). So a divergent behavior may also be found in this case. Finally, we would like to remark that for this general Hamiltonian (equation \ref{eq:Hamedge}) it is also possible to implement the reconstruction algorithm and follow the deformations and rotations of polyhedra with different number of edges in an analogous way as done in section \ref{subsec:generalrandom} for the LQG Hamiltonian.

\section*{Conclusions}

The parametrization of the loop gravity phase space in terms of spinors and its subsequent quantization \cite{Livine:2011gp,Livine:2013zha,Dupuis:2011dyz,Borja:2010rc} has been successfully used in the past in order to study several key aspects of loop quantum gravity (LQG) and loop quantum cosmology (LQC), such as the construction of coherent states \cite{Freidel:2010tt,Calcinari:2020bft}, the formulation of dynamics within symmetry reduced models \cite{Borja:2010gn,Borja:2010rc} and the construction of classical settings for cosmological models \cite{Livine:2011up}.
This spinorial formalism for LQG provides an interesting geometrical perspective. The imposition of the closure constraint allows the association of a convex polyhedron of $N$ faces to a collection of $N$ closed spinors (using the Minkowski theorem). This geometrical interpretation of the spinors has been extensively explored and its results have been found to be specially appealing \cite{Freidel:2010aq,Charles:2015lva,Goeller:2018jaj}.

Following the work done in \cite{Borja:2010rc}, we have considered a simple model given by a graph with 2 vertices linked by $N$ edges. Nevertheless, in the present paper we have studied this model in the general case, without restricting ourselves to the symmetry reduced sector (homogeneous and isotropic sector) explored there. 

In order to study the classical dynamics, we have first considered the so-called LQG Hamiltonian (equation \ref{eq:LQGham}) that was previously proposed in \cite{Borja:2010rc}. The building observables of this Hamiltonian satisfy a closed algebra with the Poisson brackets and it seems the most natural Hamiltonian to study the dynamics of the system, although alternative normalizations (still commuting with the closure and matching constraints) are also interesting and have been considered. In particular, the Hamiltonian corresponding to a lattice gauge theory (using the holonomies around closed loops) provides a useful perspective.

The equations of motion and the dynamics of the polyhedra associated with the classical spinors have been deeply explored for the LQG Hamiltonian. 
The evolution in this general case (out of the homogeneous and isotropic sector) is much more intricate than in the reduced case, given that the equations of motion constitute  a system of coupled non-linear differential equations for the components of the spinors.

We have solved the equations of motion for the simplest case of only $2$ edges. The results that we have obtained show the emergence of different regimes (oscillatory or divergent) depending only on the values of the coupling constants of the Hamiltonian. In fact, we obtain the same regimes that appeared analytically in the reduced case studied in \cite{Borja:2010gn,Borja:2010rc}, where interesting cosmological analogies and interpretations were also explored.

Furthermore, we have considered the important case of the tetrahedron (4 edges). In the first place, with a parametrization of the spinors corresponding to the isosceles tetrahedron \cite{Goeller:2018jaj} and, later on, for the general case with random spinors. Notice that within the LQG theory the tetrahedron is the simplest possible model with non-zero volume, so this case is specially relevant. The evolution of the spinors, areas of the faces, volume and the quadrupole moment for this case has been studied, obtaining again the same regimes in terms of the coupling constants of the Hamiltonian.

Finally, the most general case with an arbitrary number of edges and with random initial configurations of the spinors has been explored. In order to obtain closed initial configurations, we have used the method described in appendix \ref{app:toolbox}, that is based on previous techniques developed in \cite{Livine:2012cv,Livine:2013tsa}. At this point, in order to gain intuition about the evolution of the discrete geometry, we needed to reconstruct the polyhedron corresponding to a certain configuration of spinors. For the case of the tetrahedron, the reconstruction could be developed using the simple method given in \cite{Charles:2015lva}. Nevertheless, for general polyhedra the reconstruction algorithm is more complex \cite{Bianchi:2010gc,Sellaroli:2017wwc}. Using Sellaroli's algorithm \cite{Sellaroli:2017wwc} for each step of the evolution, we have been able to study the evolution of the polyhedra, as well as their areas and volumes.
Additionally, we have checked that we may use the eigenvalues of the quadrupole moment of the polyhedron to sketch the behavior of the volume. By doing so, we do not need to use the reconstruction algorithm for each time step and, thus, the computational time decreases drastically. Interestingly, in this general case the same oscillatory and divergent regimes are obtained (marked by the same relation between the coupling constants). 

We also explored the dynamics given by a much more general Hamiltonian with edge dependent coupling constants. We solved and plotted the evolution for random tetrahedra. However, no regimes were found.

To summarize, we have used the spinorial formalism for LQG to explore the 2-vertex model, which was already shown to be useful in order to study non-trivial dynamics within the LQG framework and where interesting cosmological implications were found. In this case, we have studied the general phase space of these models (extending the results found in \cite{Borja:2010rc} for the homogeneous and isotropic sector). Moreover, making use of Sellaroli's reconstruction algorithm, we analyzed the evolution of the polyhedra associated with randomly chosen initial spinors. Remarkably, for the LQG Hamiltonian considered, we obtained a universal behavior of the evolution that showed either an oscillatory or a divergent regime depending only on the sign of a combination of the coupling constants, 
and it was thus completely independent of the initial configuration of the spinors or the number of edges, i.e., independent of the initial polyhedron considered. Given the cosmological implications of this models \cite{Borja:2010gn,Borja:2010rc,Livine:2011up}, this fact opens an interesting path to explore more general cosmological models within the LQG theory.

\section*{Acknowledgments}

IG acknowledges financial support from the Basque Government Grant No. IT956-16 and from the Grant FIS2017-85076-P, funded by MCIN/AEI/10.13039/501100011033 and by ``ERDF A way of making Europe''.

We are also very grateful to Pietro Dona for his explanations on the reconstruction of polyhedra from intertwiner data and for pointing us to the reconstruction algorithm and code given in reference \cite{Sellaroli:2017wwc}.

\appendix

\section{The spinorial toolbox}
\label{app:toolbox}

There are plenty of mathematical results on the ensemble of polyhedra described by the face normals and averages over this ensemble \cite{Livine:2012cv,Livine:2013tsa}. We will review in this appendix some of these results and we will propose a method in order to generate random spinors for each of the vertices of the 2-vertex model, satisfying both the closure and matching constraints.

\subsection{Closing a randomly generated collection of spinors}

In the first place, we will generate a collection of 3d vectors with random directions on the sphere and with Gaussian distribution on their norms. Then, we consider the associated spinors to each 3d vector of the collection $\{z_{i}\}_{i=1..N}$, which do not necessarily satisfy the closure constraint, i.e. the 2$\times$2-matrix\, $C=\sum_{i}|z_{i}\ra\la z_{i}| $ is not proportional to the identity $\id_{2}$. 

Since $C$ is a positive Hermitian matrix, we can diagonalize it and take its square-root:
\be
C=\sum_{i}|z_{i}\ra\la z_{i}| =
\rho\Lambda \Lambda^{\dagger}\,,\qquad
\textrm{with}\quad
\rho=\det C>0\,,\quad
\Lambda\in\SL(2,\C)\,.
\label{eq:equality}
\ee
Such a matrix $\Lambda$ is not unique and we can compose it with an arbitrary $\SU(2)$ matrix, $\Lambda \rightarrow \Lambda g$ with $g\in\SU(2)$, without changing the equality above \eqref{eq:equality}. This allows for instance to fix $\Lambda$ to be a pure boost, i.e. $\Lambda=\Lambda^{\dagger}$.

One can then act with the inverse Lorentz transformation (whether we fixed $\Lambda$ to be a pure boost or not) and define a modified collection of spinors $\tz_{i}$:
\be
|\tz_{i}\ra=\Lambda^{-1}|z_{i}\ra\,,\qquad
\tilde{C}=\sum_{i}|\tz_{i}\ra\la \tz_{i}| =\Lambda^{-1}C\Lambda=\rho\id_2\,.
\ee
This new set of spinors is closed by construction. It is even the unique closed collection of spinors, which can be obtained from the original spinors by the action of $\SL(2,\C)$. The total area $\tr\, C$, as well as the individual areas $\la z_{i}|z_{i}\ra$ are obviously not conserved by this action. Nevertheless, as proved in \cite{Livine:2013tsa}, the $\SL(2,\C)$ orbits are at constant $[z_{i}|z_{j}\ra$ for every pair $(i,j)$.

\subsection{Deforming a collection of spinor at fixed areas}

Once we have obtained a collection of closed random spinors for one of the vertices of the model, we would like to obtain a different closed collection of spinors for the other vertex that satisfies the matching constraint. Therefore, the idea is to deform the original closed collection of spinors without changing the individual areas $\la z_{i}|z_{i}\ra$. 

Previous work \cite{Freidel:2009ck,Borja:2010gn,Livine:2013tsa} showed that one can explore the whole space of closed collection of spinors at fixed total area $\sum_{i }\la z_{i}|z_{i}\ra$ by $\U(N)$ transformations generated by the (Poisson flows of the) observables $E_{ij}=\la z_{i}|z_{j}\ra$. Those transformations actually trade the area of the $i$-th face for the area of the $j$-th face, and thus clearly change the individual areas. Nevertheless, we can use them to cook up a flow that modifies the face normals without changing the face areas. We simply use:
\be
S_{ij}=E_{ij}E_{ji}=|\la z_{i}|z_{j}\ra|^{2}
\,,\qquad
S_{ij}=S_{ji}
\,.
\ee
This is basically the scalar product $\vX_{i}\cdot\vX_{j}$. It clearly commutes with the individual areas and with the total closure vector (since it is $\SU(2)$-invariant):
\be
\{S_{ij},\la z_{k}|z_{k}\ra\}=0\,,\qquad
\{S_{ij},\sum_{k}\la z_{k}|\vsigma|z_{k}\ra\}=0\,.
\ee
It nevertheless modifies the flux vectors:
\be
\begin{array}{lcr}
\{S_{ij},|z_{i}\ra\}
&=&
i\la z_{j}|z_{i}\ra\,|z_{j}\ra \,,
\\
\{S_{ij},\la z_{i}|\}
&=&
-i\la z_{i}|z_{j}\ra\,\la z_{j}|\,,
\end{array}
\qquad
\{S_{ij},\la z_{i}|\vsigma|z_{i}\ra\}
=
i
\,\Big{[}
\la z_{j}|z_{i}\ra\,\la z_{i}|\vsigma|z_{j}\ra
-
\la z_{i}|z_{j}\ra\,\la z_{j}|\vsigma|z_{i}\ra
\Big{]}
\,.
\ee
This generically does not vanish. For instance, let us choose $z_{i}=(1,0)$, then the right-hand side is easily evaluated to:
\be
i\Big{[}
\la z_{j}|z_{i}\ra\,\la z_{i}|\vsigma|z_{j}\ra
-
\la z_{i}|z_{j}\ra\,\la z_{j}|\vsigma|z_{i}\ra
\Big{]}
=
\left(
\begin{array}{c}
i(\bz^{0}_{j}z^{1}_{j}-z^{0}_{j}\bz^{1}_{j}) \\
(\bz^{0}_{j}z^{1}_{j}+z^{0}_{j}\bz^{1}_{j}) \\
0
\end{array}
\right)\,.
\ee
Another way to see that this flow does not vanish is to compute directly the Poisson bracket of the scalar product $\vX_{i}\cdot\vX_{j}$ with $\vX_{i}$:
\be
\{\vX_{i}\cdot\vX_{j}\,,\,\vX_{i}\}
\propto
\vX_{i}\w \vX_{j}
\,.
\ee
Thus, in order to deform a closed spinor configuration while preserving the individual areas, one can simply let the spinors evolve along the Hamiltonian flow generated by $S_{ij}$, given by the non-linear coupled evolution equations:
\be
\pp_{t_{ij}}|z_{i}\ra=-\{S_{ij},|z_{i}\ra\}=-i\la z_{j}|z_{i}\ra\,|z_{j}\ra
\,,\qquad
\pp_{t_{ij}}|z_{j}\ra=-i\la z_{i}|z_{j}\ra\,|z_{i}\ra
\,,\qquad
\pp_{t_{ij}}|z_{k}\ra=0\quad\forall k\ne i,j
\,.
\ee

\medskip

Now, if we would like to generate a (seemingly) almost random closed spinor configuration from a given original spinor configuration with the same face areas, one could let the spinors evolve during a random amount of time $t_{i_{1}i_{2}}$ along the flow generated by $S_{i_{1}i_{2}}$ for a randomly chosen pair $(i_{1},i_{2})$, then choose a random index $i_{3}$ and let the spinors evolve during a random amount of time $t_{i_{2}i_{3}}$ along the flow generated by $S_{i_{2}i_{3}}$, and so on.

\section{Method for the statistical study of the parameter space}\label{appNumplots}

The cases considered along this paper lay out of the homogeneous and isotropic sector (symmetry reduced sector) of the 2-vertex model studied in \cite{Borja:2010gn,Borja:2010rc}. In that case, it was possible to solve analytically the equations of motion and study the different regimes in the parameter space. Nevertheless, for the general cases considered here, even though we obtained the exact analytical equations of motion for our models, these are solved numerically. Therefore, the search for the sectors of the parameter space leading to the oscillatory or divergent regimes has been done by resorting to numerical methods and analyzing a large sample of the parameter space.

\begin{figure}[htb]
\begin{center}
\includegraphics[width=0.95\textwidth]{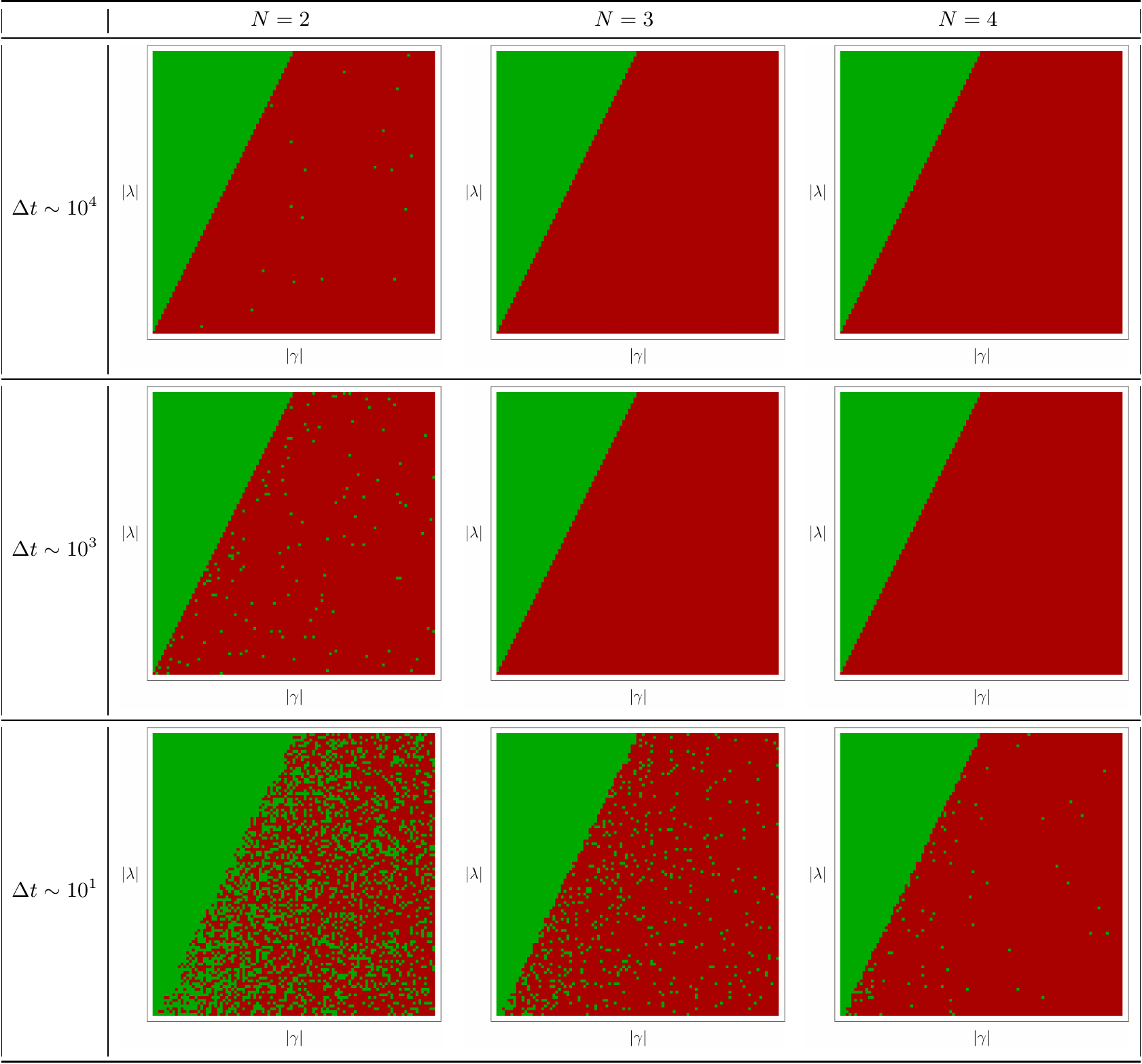}
  \caption{In this table of figures the different regimes for $N=2$ (left column), $N=3$ (middle column) and $N=4$ (right column) are represented.
    Green colored pixels represent the values of $\lambda$ and $\gamma$ for which the evolution does not diverge, whereas the red color represents the divergent regimes. The red pixels are obtained by looking whether an error is raised in the code, which only will appear if a divergence is found when solving the system of differential equations. Each pixel corresponds to a randomly chosen set of initial values for the spinors, which satisfy both the area-matching and closure constraints (each plot contains $10^4$ pixels). The different rows represent the different time intervals considered in order to find a divergence. We can observe that for $N=2$, we need large values of $\Delta t$ to see the divergence. However, for $N=3$ and $N=4$, we observe that the divergence appears earlier. All the plots reflect that the oscillatory and divergent regimes appear depending on the value of $\text{Sign}(\lambda^2-4|\gamma|^2)$ (represented by the border line between the red and green sectors).}
    \label{tbl:table_color_regimes}
\end{center}
\end{figure}

For the models with 2 and 4 edges, studied in sections \ref{subsec:2edgesHLQG} and \ref{sec:tetrahedron} respectively, we have considered a large number of initial values for the spinors and coupling constants. Systematically, we have changed the initial values of the spinors and phases, and we have chosen different values of the coupling constants $\lambda$ and $\gamma$ trying to span as many scenarios as possible. We have crossed real positive and real negative values, as well as imaginary numbers with positive and negative components. On the other hand, setting the initial conditions for the spinors, we have changed the values of $\lambda$ and $\gamma$. For every set of initial spinors, both for the $N=2$ and $N=4$ cases, we have obtained the same trends as shown in figure \ref{tbl:table_color_regimes} (the figure is done for the general random case). However, taking other initial values for the spinors led to equal results, proving thus the consistency of our analysis.

For the more general models studied in section \ref{subsec:generalrandom}, we have also obtained these kind of results. In these cases, however, since we have developed an algorithm (appendix \ref{app:toolbox}) to build the initial configurations which already satisfy the matching and closure constraints, we only needed to set the coupling constants. An interesting result that we have observed (apart from the appearance of these regimes even for the general random cases) is that the divergent regimes diverge earlier for higher number of edges (see figure \ref{tbl:table_color_regimes}).

Finally, we have also studied in the same systematic way the cases where the coupling constants of the Hamiltonian depend on the edges ($\lambda_{ij}$ and $\gamma_{ij}$). Nevertheless, as commented in section \ref{sec:edge-dep}, we have not found any trend which describes the dependence of the regimes on the coupling constants and initial values of the spinors.

\providecommand{\href}[2]{#2}\begingroup\raggedright\endgroup

\end{document}